\documentclass[prd,preprint,superscriptaddress,preprintnumbers,eqsecnum,showpacs,nofootinbib,nobibnotes]{revtex4}
\usepackage{amsfonts,amsmath,amssymb,bm,natbib}
\usepackage{graphicx} 
\usepackage{mathtools}
\usepackage{boondox-cal}
\usepackage{xcolor}

\usepackage[utf8]{inputenc}
\usepackage{slashed}

\newcommand{\be}{\begin{equation}}
\newcommand{\bea}{\begin{eqnarray}}
\newcommand{\ba}{\begin{align}}
\newcommand{\ee}{\end{equation}}
\newcommand{\eea}{\end{eqnarray}}
\newcommand{\ea}{\end{align}}

\def\1eq#1{Eq.~(\ref{#1})}

\def\2eqs#1#2{Eqs.~(\ref{#1}) and~(\ref{#2})}
\def\3eqs#1#2#3{Eqs.~(\ref{#1}),~(\ref{#2}) and~(\ref{#3})}
\def\4eqs#1#2#3#4{Eqs.~(\ref{#1}),~(\ref{#2}),~(\ref{#3}) and~(\ref{#4})}
\def\noeq#1{(\ref{#1})}

\def\tbarc{\bar{\mathcal{c}}^*}

\def\s#1{{\scriptscriptstyle #1}}

\def\G{\Gamma}

\def\T{T_1}

\def\s{\mathcal{s}}

\def\hphi0{{\hat\phi}_0}

\def\d{\!\mathrm{d}^4x\,}
\def\dl{\delta_{\xi0}}
\def\div{\frac1\epsilon}

\def\cfxt#1{\theta_{#1}}
\def\cfct#1{\rho_{#1}}
\def\cfgi#1{\lambda_{#1}}
\def\cfps#1{\vartheta_{#1}}
\def\ff#1#2{\gamma^{#1}_{#2}}

\makeatletter
\def\user@resume{resume}
\def\user@intermezzo{intermezzo}
\newcounter{previousequation}
\newcounter{lastsubequation}
\newcounter{savedparentequation}
\renewenvironment{subequations}[1][]{%
      \def\user@decides{#1}%
      \setcounter{previousequation}{\value{equation}}%
      \ifx\user@decides\user@resume 
           \setcounter{equation}{\value{savedparentequation}}%
      \else  
      \ifx\user@decides\user@intermezzo
           \refstepcounter{equation}%
      \else
           \setcounter{lastsubequation}{0}%
           \refstepcounter{equation}%
      \fi\fi
      \protected@edef\theHparentequation{%
          \@ifundefined {theHequation}\theequation \theHequation}%
      \protected@edef\theparentequation{\theequation}%
      \setcounter{parentequation}{\value{equation}}%
      \ifx\user@decides\user@resume 
           \setcounter{equation}{\value{lastsubequation}}%
         \else
           \setcounter{equation}{0}%
      \fi
      \def\theequation  {\theparentequation  \alph{equation}}%
      \def\theHequation {\theHparentequation \alph{equation}}%
      \ignorespaces
}{%
  \ifx\user@decides\user@resume
       \setcounter{lastsubequation}{\value{equation}}%
       \setcounter{equation}{\value{previousequation}}%
  \else
  \ifx\user@decides\user@intermezzo
       \setcounter{equation}{\value{parentequation}}%
  \else
       \setcounter{lastsubequation}{\value{equation}}%
       \setcounter{savedparentequation}{\value{parentequation}}%
       \setcounter{equation}{\value{parentequation}}%
  \fi\fi
  \ignorespacesafterend
}
\makeatother

\begin{document}

\title{
Off-shell renormalization\\ in the presence of dimension~6 derivative operators.\\ 
II. UV coefficients}

\date{August 23, 2019}

\author{D. Binosi}
\email{binosi@ectstar.eu}
\affiliation{European Centre for Theoretical Studies in Nuclear Physics
and Related Areas (ECT*) and Fondazione Bruno Kessler, Villa Tambosi, Strada delle Tabarelle 286, I-38123 Villazzano (TN), Italy}

\author{A. Quadri}
\email{andrea.quadri@mi.infn.it}
\affiliation{INFN, Sezione di Milano, via Celoria 16, I-20133 Milano, Italy}

\begin{abstract}
\noindent
The full off-shell one loop renormalization 
for all divergent amplitudes up to dimension 6 in the
Abelian Higgs-Kibble model, supplemented with a maximally power counting violating higher-dimensional gauge-invariant derivative interaction $\sim g ~ \phi^\dagger \phi (D^\mu \phi)^\dagger D_\mu \phi$,
is presented.
This allows one to perform the complete renormalization of radiatively generated dimension 6 operators in the model at hand.
We describe in details the technical tools required 
in order to disentangle the contribution to UV divergences parameterized by
(generalized) non-polynomial field redefinitions.
We also discuss how to extract
the dependence of the $\beta$-function coefficients 
on the non-renormalizable coupling $g$ in one loop approximation, as well as the cohomological
techniques (contractible pairs) required to
efficiently separate the mixing of contributions associated to
different higher-dimensional operators in a
spontaneously broken effective field theory.

\end{abstract}

\pacs{
11.10.Gh, 
12.60.-i,  
12.60.Fr 
}

\maketitle

\section{Introduction}

In this paper we continue the study of the off-shell renormalization of the Abelian Higgs-Kibble model supplemented by the maximally power counting  violating dimension 6 operator $\phi^\dagger \phi (D^\mu \phi)^\dagger D_\mu \phi$. In particular, we will show here how to evaluate the one-loop divergent coefficients associated to {\it all} dimension 6 operators which are radiatively generated.

The general aspects of the formalism needed to achieve this result have been explained in details in~\cite{BQ:2019a}, to which we refer the reader for a thorough exposition of the technical tools required within the Algebraic Renormalization approach to the problem~\cite{Piguet:1995er,Ferrari:1999nj,Grassi:1999tp,Grassi:2001zz,Quadri:2003ui,Quadri:2003pq,Quadri:2005pv,Bettinelli:2007tq,Bettinelli:2007cy,Bettinelli:2008ey,Bettinelli:2008qn,Anselmi:2012qy,Anselmi:2012jt,Anselmi:2012aq} we use. 

The present paper describes in a self-contained way the procedure developed in~\cite{BQ:2019a} from an operational point of view. In particular we show how to disentangle the contributions to UV divergences parameterzied by unphysical (generalized) non-polynomial field redefinitions from those 
associated to the renormalization of physical gauge-invariant operators in the evaluation of one-loop $\beta$-functions.

To systematically compute the (one-loop) UV coefficients in spontaneously broken effective field theories possessing (dimension 6) derivative operators, it is convenient to first renormalize an associated auxiliary model, the so-called $X$-theory, which is obtained by describing the scalar physical degree of freedom in terms of the gauge-invariant field coordinate
\begin{align}
	v X_2 \sim  \phi^\dagger \phi - \frac{v^2}{2},
	\label{constr.intr}
\end{align}  
$v$ being the vacuum expectation value of the Higgs scalar $\phi$.

Then, in the $X$-theory all higher dimensional operators in the classical action are required to vanish at $X_2=0$. Thus, the operator $\frac{g}{v\Lambda} \phi^\dagger \phi (D^\mu \phi) D_\mu \phi$ (with the energy scale $\Lambda$ much higher than the electroweak scale $v$) will be expressed as $\frac{g}{\Lambda}X_2 (D^\mu \phi) D_\mu \phi$; going on-shell with the field $X_2$ and an additional Lagrange multiplier $X_1$ enforcing algebraically the constraint in~\1eq{constr.intr}, we get back the original operator. Two external sources are then required in order to formulate in a mathematically consistent way the $X$-theory~\cite{BQ:2019a}: one is coupled to the constraint $v X_2 -  \phi^\dagger \phi - \frac{v^2}{2}$ and is denoted by $\bar c^*$; the second, called $T_1$, is required to close the algebra of operators, implementing the $X_2$-equation of motion at the quantum level.

The important point is that, unlike in the ordinary formalism, in the $X$-theory all 1-PI amplitudes, with the exception of those involving insertions of the $T_1$ source, exhibit a manifest weak power-counting~\cite{Ferrari:2005va}: only a finite number of divergent amplitudes exist at each loop order (although increasing with the loop number, as expected in a general effective field theory setting). As for $T_1$-dependent amplitudes, they can be recovered by resumming the $T_1$-insertions on the Green's functions at $T_1=0$, which, sometimes, can be even done in a closed form.

Once the renormalization of the $X$-theory is achieved, one goes on-shell with the $X_{1}$ and $X_2$ fields, which amounts to a suitable mapping of the sources $\bar c^*$ and $T_1$ onto operators depending on $\phi$ and its covariant derivatives. Then,  one can immediately read off the UV coefficients of the higher dimensional gauge-invariant operators in the target theory, as now everything is expressed in terms of the original $\phi$ field. We hasten to emphasize that since we are working off-shell the effects of generalized field redefinitions, that are present already at one-loop order, and are not even polynomial for the model at hand~\cite{BQ:2019a}, need to be correctly accounted for. This is automatically done through the cohomologically trivial invariants of the $X$-theory. In fact, as we will show, the associated coefficients are gauge-dependent (as we will explicitly check by evaluating all the coefficients both in Feynman and Landau gauge), being instrumental in ensuring crucial cancellations leading to the gauge-independence of the coefficients associated to gauge-invariant operators. 
Notice in fact that since the ensuing analysis is based  on cohomological results valid for anomaly-free gauge theories, the computational approach presented here can be readily extended to the electroweak gauge group $\rm SU(2) \times U(1)$ and,  more generally, to any non-anomalous non-Abelian gauge group.

The paper is organized as follows. Our notations and conventions are described in Sect.~\ref{sec.not}.  After providing in Sect.~\ref{sec.map} a brief reminder on the structure of the mapping to the target theory, we proceed to evaluate the coefficients of the cohomologically trivial invariants relevant for dimension 6 operators in Sect.~\ref{sec:ct}. Sect.~\ref{sec:ps}, \ref{sec:ms} and~\ref{sec:gi} are then devoted to the evaluation of the coefficients of the three classes of gauge invariant operators appearing in the theory: those only depending on the external sources, those mixing external sources and fields and those that only depend on the fields. We finally apply the mapping to the target theory in Sect.~\ref{sec:map} thereby computing the coefficients of all the UV divergent operators up to dimension~$6$ in the original (target) theory. This allows us to construct (Sect.~\ref{sec.beta}) the full $\beta$ functions of the theory. Our conclusions and outlook are presented in Sect.~\ref{sec:concl}. The paper ends with two appendices: Appendix~\ref{app:list} contains the list of all the independent invariants needed for renormalizing the theory, while the relevant $X$-theory divergent one-loop amplitudes up to dimension $6$ are given in Appendix~\ref{app:UVdivamp}.

\section{Notations and setup}\label{sec.not}


The tree-level vertex functional in the $X$-formalism can be written as~\cite{BQ:2019a}
\begin{align}
	\G^{(0)} & = 
	   \int \!\mathrm{d}^4x \, \Big [ -\frac{1}{4} F^{\mu\nu} F_{\mu\nu} + (D^\mu \phi)^\dagger (D_\mu \phi) - \frac{M^2-m^2}{2} X_2^2 - \frac{m^2}{2v^2} \Big ( \phi^\dagger \phi - \frac{v^2}{2} \Big )^2 \nonumber \\
	& - \bar c (\square + m^2) c + \frac{1}{v} (X_1 + X_2) (\square + m^2) \Big ( \phi^\dagger \phi - \frac{v^2}{2} - v X_2 \Big ) \nonumber \\
	&  + \frac{g}{\Lambda} X_2 (D^\mu \phi)^\dagger (D_\mu \phi)  + \T(D^\mu \phi)^\dagger (D_\mu \phi) \nonumber \\
	& + \frac{b^2}{2\xi} -  b \Big ( \partial A + \frac{e v}{\xi} \chi \Big ) + \bar{\omega}\Big ( \square \omega + \frac{e^2 v}{\xi} (\sigma + v) \omega\Big ) \nonumber \\
	&  + \bar c^* \Big ( \phi^\dagger \phi - \frac{v^2}{2} - v X_2 \Big ) + \sigma^* (-e \omega \chi) + \chi^* e \omega (\sigma + v) \Big ].
	\label{tree.level}
\end{align}
The first line is the action
of the Abelian Higgs-Kibble model in the $X$-formalism.
Besides the usual scalar field
$\phi \equiv \frac{1}{\sqrt{2}} (\sigma + v + i \chi)$, with $v$ its
vacuum expectation value (v.e.v.), one also adds 
a singlet field $X_2$ that
provides a gauge-invariant
parameterization of the
physical scalar mode.
When going on-shell with the
field $X_1$, that plays the
role of a Lagrange multiplier, one recovers
the constraint\footnote{
Going on-shell with $X_1$
yields 
the condition
\begin{align}
    (\square + m^2) \Big (
    \phi^\dagger \phi - \frac{v^2}{2} - v X_2 \Big ) = 0 \, ,
\end{align}
so that the most general solution is
$X_2 = \frac{1}{v} \Big (
    \phi^\dagger \phi - \frac{v^2}{2} \Big ) + \eta,$ $\eta$ being a
    scalar field of mass $m$.
However in perturbation theory
the correlators of the mode $\eta$ with any gauge-invariant operators vanish~\cite{Binosi:2019olm}, so that one can safely set $\eta =0$.
}
$X_2 \sim \frac{1}{v} ( \phi^\dagger \phi - v^2/2)$.
Inserting the latter back into the first line
of Eq.(\ref{tree.level}), the $m^2$-term cancels out and one is left with the 
usual Higgs quartic potential
with coefficient $\sim M^2/2v^2$.

Hence, Green's functions in the target theory have to be $m^2$-independent, a fact that provides a very strong check of the computations, due to the ubiquitous presence of~$m^2$ both in Feynman amplitudes and invariants.

The $X_{1,2}$-system comes together with a 
{\em constraint} BRST symmetry, ensuring that the number of 
physical degrees of freedom in the scalar sector remains 
unchanged in the $X$-formalism w.r.t. the standard formulation relying only on the field $\phi$~\cite{Quadri:2006hr,Quadri:2016wwl}.
More precisely, the vertex functional (\ref{tree.level}) is invariant
under the following BRST symmetry:
\begin{align}
	\s X_1 = v c; \, \quad \s \phi = \s X_2 = \s c  = 0; \quad \s \bar c = \phi^\dagger\phi - \frac{v^2}{2} - v X_2 \, . 
	\label{u1.brst}
\end{align}
The associated ghost and antighost fields $c, \bar c$ are free. 
The constraint BRST differential $\s$ anticommutes with 
the {\it gauge group} BRST symmetry of the classical action after the gauge-fixing introduced in the fourth line of
Eq.(\ref{tree.level}):
\begin{eqnarray}
	s A_\mu = \partial_\mu \omega \, ;  \qquad s \omega = 0 \, ; \qquad  s \bar{\omega} = b \, ; \qquad s b =0 \, ; \quad s \phi = i e \omega \phi.  
\end{eqnarray}
Here $\omega$ ($\bar \omega$) is the U(1) ghost (antighost); the latter field is paired into a BRST doublet with the Lagrange multiplier field $b$, enforcing the $R_\xi$ gauge-fixing condition
\begin{align}
    {\cal F}_\xi = \partial A + \frac{ev}{\xi} \chi.
    \label{g.f.cond}
\end{align}
The two BRST symmetries can both
be lifted to the corresponding
Slavnov-Taylor identities at the quantum level, provided one 
introduces the antifields,
{\it i.e.}, the external sources
coupled to the relevant
BRST transformation that are non-linear in the quantized fields.
The antifield couplings are displayed
in the last line of Eq.(\ref{tree.level}).
Then the ST identity for the
constraint BRST symmetry is
\begin{align}
    {\cal S}_{\scriptscriptstyle{C}}(\G) \equiv \int \!\mathrm{d}^4 x \, \Big [ v c \frac{\delta \G}{\delta X_1} 
 + \frac{\delta \G}{\delta \bar c^*}\frac{\delta \G}{\delta \bar c} \Big ] = 
 \int \!\mathrm{d}^4 x \, \Big [ v c \frac{\delta \G}{\delta X_1} 
 -(\square + m^2) c \frac{\delta \G}{\delta \bar c^*} \Big ] = 0,
 \label{sti.c} 
\end{align}
where in the latter equality we have used the fact that both the ghost $c$ and the antighost $\bar c$ are free:
\begin{align}
    \frac{\delta \G}{\delta \bar c} = -(\square + m^2) c \, \, ,
    \qquad
     \frac{\delta \G}{\delta c} = (\square + m^2) \bar c \, .    
\end{align}
Hence Eq.(\ref{sti.c}) reduces
to the $X_1$-equation of motion
\begin{align}
    \frac{\delta \G}{\delta X_1}=
    \frac{1}{v} (\square + m^2)
    \frac{\delta \G}{\delta \bar c^*}.
    \label{X1.eq}
\end{align}
Finally, the ST identity
(equivalently the BV master equation)
associated to the gauge group
BRST symmetry reads
\begin{align}
	& {\cal S}(\G)  = \int \mathrm{d}^4x \, \Big [ 
	\partial_\mu \omega \frac{\delta \G}{\delta A_\mu} + \frac{\delta \G}{\delta \sigma^*} \frac{\delta \G}{\delta \sigma}  + \frac{\delta \G}{\delta \chi^*} \frac{\delta \G}{\delta \chi} 
	+ b \frac{\delta \G}{\delta \bar \omega} \Big ] = 0. 
	\label{sti} 
\end{align}

The third line of Eq.(\ref{tree.level}) contains the derivative dim.6 operator
$$X_2 (D^\mu \phi)^\dagger D_\mu \phi \sim \Big ( \phi^\dagger \phi - \frac{v^2}{2} \Big ) (D^\mu \phi)^\dagger D_\mu \phi$$
together
with the source $T_1$ required to define the $X_2$-equation at the quantum level
in the presence of such an additional non power-counting
renormalizable interaction:
\begin{eqnarray}
	\frac{\delta \G}{\delta X_2} =  \frac{1}{v} (\square + m^2) \frac{\delta \G}{\delta \bar c^*} + \frac{g}{\Lambda} \frac{\delta \G}{\delta T_1} - (\square + m^2)X_1 - (\square + M^2) X_2 - v \bar c^* \, .
	\label{X2.eq}
\end{eqnarray}
Notice that the terms in the third line of Eq.(\ref{tree.level}) respect both BRST symmetries and thus they do not violate either the $X_1$-equation~(\ref{X1.eq}) or the ST identity~(\ref{sti}).

The set of the functional identities
holding in this theory is completed
by:
\begin{itemize}
\item The $b$-equation:
\begin{eqnarray}
	\frac{\delta \G^{(0)}}{\delta b} = \frac{b}{\xi} - \partial A - \frac{e v}{\xi} \chi ;
	\label{b.eq}
\end{eqnarray}
\item The antighost equation:
\begin{eqnarray}
	\frac{\delta \G^{(0)}}{\delta \bar \omega} = \square \omega + \frac{e v}{\xi} \frac{\delta \G^{(0)}}{\delta \chi^*} .
	\label{antigh.eq}
\end{eqnarray}   
\end{itemize}

In what follows subscripts denote functional differentiation w.r.t. fields and external sources. Moreover, if not otherwise stated, amplitudes will be denoted as, {\it e.g.}, $\G^{(1)}_{\chi\chi}$, meaning
\begin{align}
    \G^{(1)}_{\chi\chi} \equiv \left . 
    \frac{\delta^2 \G^{(1)}}{\delta \chi(-p) \delta \chi(p)} \right |_{p=0}.
\end{align}
A bar denotes the UV divergent part of the corresponding amplitude in the Laurent expansion around $\epsilon=4-D$, with $D$ the space-time dimension. Dimensional regularization is always implied, with amplitudes evaluated by means of the packages {\tt FeynArts} and {\tt FormCalc}~\cite{Hahn:2000kx,Hahn:2000jm}. As already remarked, all amplitudes will be evaluated in the Feynman ($\xi=1$, with $\xi$ the gauge fixing parameter) and Landau ($\xi=0$) gauge; this will allow to explicitly check the gauge cancellations in gauge invariant operators.

The UV divergent contributions to one-loop amplitudes form a local functional (in the sense of formal power series) aptly denoted by $\overline{\G}^{(1)}$. In particular,~$\overline{\G}^{(1)}$ belongs to the kernel of ${\cal S}_0$ i.e.
\begin{align}
    {\cal S}_0(\overline{\G}^{(1)}) =0,
    \label{uv.div.1loop.st}
\end{align}
where ${\cal S}_0$ is the linearized ST operator
\begin{align}
	{\cal S}_0  (\overline{\G}^{(1)}) & = \int \!\mathrm{d}^4 x \, \Big [ \partial_\mu \omega \frac{\delta \overline{\G}^{(1)}}{\delta A_\mu}  + 
	e\omega(\sigma+v)\frac{\delta \overline{\G}^{(1)}}{\delta \chi}  
	-e\omega\chi\frac{\delta \overline{\G}^{(1)}}{\delta \sigma}
+ b \frac{\delta \overline{\G}^{(1)}}{\delta \bar \omega}  \nonumber \\
&  + \frac{\delta \G^{(0)}}{\delta \sigma} \frac{\delta \overline{\G}^{(1)}}{\delta \sigma^*} + \frac{\delta \G^{(0)}}{\delta \chi} \frac{\delta \overline{\G}^{(1)}}{\delta \chi^*} \Big ] \nonumber \\
& = s \overline{\G}^{(1)} + \int \!\mathrm{d}^4 x \, \Big [ \frac{\delta \G^{(0)}}{\delta \sigma} \frac{\delta \overline{\G}^{(1)}}{\delta \sigma^*} + \frac{\delta \G^{(0)}}{\delta \chi} \frac{\delta \overline{\G}^{(1)}}{\delta \chi^*}  \Big ],
\label{S0}
\end{align}
which acts as the BRST differential $s$ on the fields of the theory while mapping the antifields into the classical equations of motion of their corresponding fields. Then, the nilpotency of ${\cal S}_0$ ensures that $\overline{\G}^{(1)}$ is the sum of a gauge-invariant functional ${\overline{{\cal I}}}^{(1)}$ and a cohomologically trivial contribution ${\cal S}_0(\overline{Y}^{(1)})$:
\begin{align}
    \overline{\G}^{(1)} =
    {\overline{{\cal I}}}^{(1)}_\mathrm{gi} + {\cal S}_0(\overline{Y}^{(1)}) .
    \label{uv.div.1loop.vf}
\end{align}

\section{\label{sec.map}Mapping on the external sources}

As a result of the previous Section, we only need to determine the invariants contributing to ${\overline{{\cal I}}}^{(1)}_\mathrm{gi}$ and $\overline{Y}^{(1)}$
that will induce in the target theory operators of dimension less or equal to $6$.

To that end we first need to consider how the mapping affects the external sources $\bar c^*,T_1$.

The $X_1$- and $X_2$-equations~\noeq{X1.eq} and~\noeq{X2.eq} at loop order 
$n \geq 1$ for $\G^{(n)}$ read
\begin{align}
	\frac{\delta \G^{(n)}}{\delta X_1} &= \frac{1}{v} (\square + m^2) \frac{\delta \G^{(n)}}{\delta \bar c^*};& 
	\frac{\delta \G^{(n)}}{\delta X_2} &= \frac{1}{v} (\square + m^2) \frac{\delta \G^{(n)}}{\delta \bar c^*} +
	\frac{g}{\Lambda} \frac{\delta \G^{(n)}}{\delta T_1},
	\label{X.eqs.loops}
\end{align}
thus implying that the whole dependence on $X_1$ and $X_2$ can only arise through the combinations 
\begin{align}
	&\tbarc = \bar c^* + \frac{1}{v} (\square + m^2) (X_1+ X_2);&  &{\cal T}_1 = T_1 + \frac{g}{\Lambda} X_2.&
	\label{X2.subst}
\end{align}
In particular, \1eq{X.eqs.loops} states that the 1-PI amplitudes involving at least one $X_1$ or $X_2$ external legs are uniquely fixed in terms of amplitudes involving neither $X_1$ or $X_2$.

We now turn to the
analysis of how the 
right-hand side of Eqs.(\ref{X2.subst})
is transformed under the mapping.
For that purpose we need to impose 
the equations of motion for
$X_{1,2}$. 
At the one-loop level, we
can restrict to tree-level equations of motion for these fields. 
As already discussed, the $X_1$-equation of motion enforces the constraint $X_2 = \frac{1}{v} \Big ( \phi^\dagger \phi - \frac{v^2}{2} \Big )$. Once one takes into account this constraint, the $X_2$-equation of motion in turn yields 
\begin{align}
    (\square + m^2) (X_1 + X_2) = - (M^2 - m^2) X_2 + \frac{g}{\Lambda} (D^\mu \phi)^\dagger D_\mu \phi - v \bar c^*. 
    \label{X2.EOM}
\end{align}
By substituting the above expressions for $X_{1,2}$ into the replacement rules~\noeq{X2.subst}
we arrive at the sought-for mapping transformation (at zero external sources):
\begin{align}
 	& \tbarc \rightarrow - \frac{(M^2 - m^2)}{v^2} \Big ( \phi^\dagger \phi - \frac{v^2}{2} \Big ) + \frac{g}{v \Lambda} (D^\mu \phi)^\dagger D_\mu \phi;&
	& {\cal T}_1 \rightarrow  \frac{g}{v \Lambda} \Big ( \phi^\dagger \phi - \frac{v^2}{2} \Big ).
	\label{repl.fin.1}
\end{align}

Since the right-hand side of the above equation contains operators of dimension at least $2$, in order to obtain target operators of up to dimension $6$  it is clear that we need to consider amplitudes with up to $3$ external sources $\bar c^*$ and $T_1$. Equivalently, we can assign dimension $2$ to both $\bar c^*$ and  $T_1$ and use it in order to identify the mixed fields-external sources invariants that will contribute to target operators of up to dimension $6$. For instance  $\int \d \bar c^* \Big ( \phi^\dagger \phi - \frac{v^2}{2} \Big )$ would project onto
\begin{align}
\int \d \bar c^* \Big ( \phi^\dagger \phi - \frac{v^2}{2} \Big ) \rightarrow &
- \frac{(M^2 - m^2)}{v^2} \int \d \Big ( \phi^\dagger \phi - \frac{v^2}{2} \Big )^2 \nonumber \\
&+ \frac{g}{v \Lambda} \int \d  (D^\mu \phi)^\dagger D_\mu \phi\Big ( \phi^\dagger \phi - \frac{v^2}{2} \Big ),
\end{align}
whereas $\int \d \bar c^* \Big ( \phi^\dagger \phi - \frac{v^2}{2} \Big )^2$ would give rise to
\begin{align}
\int \d \bar c^* \Big ( \phi^\dagger \phi - \frac{v^2}{2} \Big )^2 \rightarrow
- \frac{(M^2 - m^2)}{v^2} \int \d \Big ( \phi^\dagger \phi - \frac{v^2}{2} \Big )^3,
\end{align}
where we have neglected the covariant kinetic term in the first term of~\1eq{repl.fin.1} since it would generate a dimension~$8$ operator.

Finally, the coefficients of the three possible types of invariants contributing to the $X$-theory functional ${\overline{{\cal I}}}^{(1)}_\mathrm{gi}$ will be indicated with $\cfgi{i}$ (combinations of the field strength, its derivatives and $\phi$ and its covariant derivatives of up to dimension $6$), $\cfxt{i}$ (combinations of external sources and fields) or $\cfps{i}$ (combinations of external sources only). The complete list of invariants is reported in Appendix~\ref{app:list}.\\

\section{\label{sec:ct}Cohomologically trivial invariants}

Before addressing the evaluation of the coefficients of the gauge invariants, it is necessary to fix the coefficients $\cfct{i}$ of the cohomologically trivial invariants contributing to ${\cal S}_0(\overline{Y}^{(1)})$. Taking into account the bounds on the dimensions, this requires to consider two invariants at $T_1=0$, namely
\begin{align}
\cfct{0}  {\cal S}_0 \! \int \d [\sigma^*(\sigma + v) + \chi^* \chi];&
    &\cfct{1}\, {\cal S}_0\! \int \d \, (\sigma^* \sigma  + \chi^* \chi).
\end{align}

\subsection{Generalized field redefinitions}

To begin with let us observe that~\1eq{S0} implies
\begin{align}
    \cfct{1}\, {\cal S}_0\! \int \d \, (\sigma^* \sigma  + \chi^* \chi) \supset - e v  \cfct{1} \int \d
    \chi^* \omega.  
    \label{rho1}
\end{align}
Therefore, the coefficient $\cfct{1}$ associated to this invariant is controlled by the single amplitude $\overline{\Gamma}^{(1)}_{\chi^* \omega}$. Indeed, \1eq{rho1} demands that
\begin{align}
     ev \cfct{1} = - \overline{\Gamma}^{(1)}_{\chi^* \omega},
\end{align}
or, using the result~\noeq{Gch*om},
\begin{align}
     \cfct{1} = \frac{M_A^2}{8 \pi^2 v^2} \div (1-\dl),
     \label{cfct.1}
\end{align}
with $\delta_{\xi0}=\delta_{00}=1$ in the Landau gauge and $\delta_{\xi0}=\delta_{10}=0$ in the Feynman gauge. Notice that this result implies that there are no {\em pure} field redefinitions in Landau gauge, {\it i.e.}, the v.e.v. renormalizes in the same way as the fields, as we will soon show.

Finally, repeated insertions of the source $T_1$ resum to 
\begin{align}
    \cfct{1} {\cal S}_0 \! \int \d \, \frac{1}{1+T_1} (\sigma^* \sigma  + \chi^* \chi). 
    \label{cfct.1.resum}
\end{align}
A comment is in order here. In the standard formalism one should consider the effect of the generalized field redefinitions in the target theory, which, as explained in~Ref.\cite{BQ:2019a}, is the one induced by~\1eq{cfct.1.resum}. This implies that the fields $\sigma$ and $\chi$ undergo the transformation
\begin{align}
    & \sigma \rightarrow \sigma  + \frac{\cfct{1}}{1+ \frac{g}{\Lambda v} \Big ( \phi^\dagger \phi - \frac{v^2}{2} \Big )} \sigma;&
    & \chi \rightarrow \chi  + \frac{\cfct{1}}{1+ \frac{g}{\Lambda v} \Big ( \phi^\dagger \phi - \frac{v^2}{2} \Big )} \chi.
\end{align}
This would be a rather involved task, which is however simplified in the approach developed here, since all the combinatorics is automatically taken into account via the renormalization of 
the $X$-theory, through the cohomologically trivial invariant~\1eq{cfct.1.resum}.

    
\subsection{Tadpoles}\label{sec:tadpoles}

The tadpoles $\overline{\G}^{(1)}_{\sigma}, \overline{\G}^{(1)}_{\bar c^*}$ allow to fix the coefficients of three invariants: 
\begin{align}
    & \cfct{0}\, {\cal S}_0 \! \int \d [\sigma^*(\sigma + v) + \chi^* \chi] + \cfgi{1} \int \d \, \Big ( \phi^\dagger \phi - \frac{v^2}{2} \Big ) 
    + \cfps{1} \int \d \bar c^*
    \nonumber \\
    & \supset \int \d \left[ ( - m^2 v \cfct{0} + v \cfgi{1} ) \sigma + ( \cfct{0} v^2 + 
    \cfps{1}
    ) \bar c^* \right ].
    \label{l1.bc}
\end{align}
Indeed, \1eq{l1.bc} gives rise to the equations
\begin{subequations}
\begin{align}
	- m^2 v \cfct{0} + v \cfgi{1} &=\overline{\Gamma}^{(1)}_{\sigma};\label{tadsa}\\
	\cfct{0} v^2 + 
	\cfps{1}
	&=\overline{\Gamma}^{(1)}_{\bar c^*}.
	\label{tadsb}
\end{align}
\end{subequations}
Direct inspection of the one-loop results~\noeq{Gc*} and~\noeq{Gs} shows that, in the Feynman gauge, it is consistent to set $\left .\cfct{0}\right |_{\xi=1}=0$, thus yielding the results
\begin{subequations}
\begin{align}
    \left.\cfgi{1}\right|_{\xi=1} & = \frac{1}{v} \left . \overline{\Gamma}^{(1)}_{\sigma} \right |_{\xi=1} = 
    \frac{1}{16 \pi^2 v^2}\div \left[ m^2 ( M^2 + M_A^2) + 2 (M^4 + 3 M_A^4) \right], \\
	\left.
	\cfps{1}
	\right|_{\xi=1} & = \left . \overline{\Gamma}^{(1)}_{\bar c^*} \right |_{\xi=1}  =  - \frac{M^2+ M_A^2 }{16 \pi^2}\div .
\end{align}	
\end{subequations}

On the other hand, since $\cfgi{1}$ must be gauge invariant, \1eq{tadsa}  implies
\begin{align}
	\cfct{0}= \frac{1}{m^2v} \Big ( v\cfgi{1}-
	\overline{\Gamma}^{(1)}_{\sigma}
	\Big ) = \frac{M_A^2}{16 v^2\pi^2} \div \dl,
    \label{cfct.0}
\end{align}
whereas~\1eq{tadsb} furnishes a consistency condition that can be easily checked. 
Notice in particular that~\1eq{tadsb} shows that 
$\cfps{1}$
is gauge independent (as it should) since the gauge dependence in $\overline{\Gamma}^{(1)}_{\bar c^*}$ is cancelled by the one in $\cfct{0}$. Finally, using~\1eq{cfct.0} and the gauge independence of 
$\cfps{1}$, \1eq{tadsb} can be recast in the form
\begin{align}
    -\frac{m^2}{v} \Big ( \left . \overline{\Gamma}^{(1)}_{\bar c^*} \right |_{\xi=0} - 
    \left . \overline{\Gamma}^{(1)}_{\bar c^*} \right |_{\xi=1} \Big ) =
    \left . \overline{\Gamma}^{(1)}_{\sigma} \right |_{\xi=0} - 
    \left . \overline{\Gamma}^{(1)}_{\sigma} \right |_{\xi=1} 
    \label{cc.bc}.
\end{align}

Next, we need to consider the insertion of one and two sources $T_1$ on tadpole amplitudes. Starting from a single insertion, the relevant projection equation becomes
\begin{align}
    & \cfct{0T_1} {\cal S}_0  \int \d  
    T_1  [\sigma^*(\sigma + v) + \chi^* \chi] + 
    \cfxt{2} \int \d   T_1 \Big ( \phi^\dagger \phi - \frac{v^2}{2} \Big ) +
    \cfps{7}
    \int \d \bar c^* T_1
    \nonumber \\
    & \supset\int \d  \Big [ ( - m^2 v \cfct{0T_1} + v \cfxt{2} ) T_1 \sigma + (v^2 \cfct{0T_1} + 
    \cfps{7}
    ) \bar c^* T_1 \Big ].
    \label{proj.1}
\end{align}
As before, one obtains two equations
\begin{subequations}
	\begin{align}
		- m^2 v \cfct{0T_1} + v \cfxt{2}&=\overline{\Gamma}^{(1)}_{T_1\sigma},\label{Ta}\\
		v^2 \cfct{0T_1} + 
		\cfps{7}
		&=\overline{\Gamma}^{(1)}_{\bar c^* T_1},\label{Tb}
	\end{align}
\end{subequations}
which is most easily solved in the Feynman gauge in which $\left.\cfct{0T_1}\right|_{\xi=1}=0$, and therefore, using the results~\noeq{GT1s} and~\noeq{Gc*T1},
\begin{subequations}
	\begin{align}
    	\left.\cfxt{2}\right|_{\xi=1} &= \frac{1}{v} \left . \overline{\Gamma}^{(1)}_{T_1\sigma} \right |_{\xi=1}
    = - \frac{1}{8 \pi^2 v^2} \Big [ m^2( M^2 + M_A^2) + 2 (M^4 - 3 M_A^4) \Big ] \div ,  \\
   	\left. 
   	\cfps{7}
   	\right|_{\xi=1} &= \left . \overline{\Gamma}^{(1)}_{\bar c^* T_1} \right |_{\xi=1} = \frac{(M^2+ M_A^2)}{8 \pi^2}  \div .
    \label{theta.2bcT1}
	\end{align}
\end{subequations}
Then, using the fact that $\theta_2$ is gauge invariant, \1eq{Ta} can be used to fix the coefficient~$\cfct{0T_1}$, obtaining 
\begin{align}
	\cfct{0T_1}&=\frac{1}{m^2v} \Big ( v\cfxt{2}-
	\overline{\Gamma}^{(1)}_{T_1\sigma}
	\Big ) = -\frac{M_A^2}{8\pi^2v^2} \div \dl,
\end{align}
which, once inserted in~\1eq{Tb} shows that 
$\cfps{7}$
is gauge invariant, thus allowing to recast the condition~\noeq{Tb} in the form
\begin{align}
    -\frac{m^2}{v} \Big ( \left . \overline{\Gamma}^{(1)}_{\bar c^*T_1} \right |_{\xi=0} - 
    \left . \overline{\Gamma}^{(1)}_{\bar c^*T_1} \right |_{\xi=1} \Big ) =
    \left . \overline{\Gamma}^{(1)}_{T_1\sigma } \right |_{\xi=0} - 
    \left . \overline{\Gamma}^{(1)}_{T_1\sigma } \right |_{\xi=1},
    \label{cc.bcT}
\end{align} 
in complete analogy with~\1eq{cc.bc}. 

Finally, for the case of two $T_1$-insertions, the relevant projection equation reads 
\begin{align}
    & \cfct{0T_1^2} \int \d 
    T_1^2 {\cal S}_0  \int \d  [\sigma^*(\sigma + v) + \chi^* \chi] + 
    \cfxt{12} \int \d  T_1^2 \Big ( \phi^\dagger \phi - \frac{v^2}{2} \Big ) +
    \frac{\cfps{11}}{2}
    \int \d  \bar c^* T_1^2
    \supset
    \nonumber \\
    & \int \d \Big [ ( - m^2 v \cfct{0T_1^2} + v \cfxt{12} ) \sigma T_1^2  + (v^2 \cfct{0T_1^2} + 
    \frac{\cfps{11}}{2}
    ) \bar c^* T_1^2 \Big ],
    \label{proj.2}
\end{align}
giving rise to the conditions
\begin{subequations}
	\begin{align}
		2 (- m^2 v \cfct{0T_1^2} + v \cfxt{12} ) &=\overline{\Gamma}^{(1)}_{\sigma T_1T_1},\label{T2a}\\ 
		2 v^2 \cfct{0T_1^2} + 
		\cfps{11} &=\overline{\Gamma}^{(1)}_{\bar c^* T_1T_1}. \label{T2b}
	\end{align}
\end{subequations}
In the Feynman gauge $\left.\cfct{0T_1^2}\right|_{\xi=1}=0$, so that, using~\2eqs{GsT1T1}{Gc*T1T1} 
\begin{subequations}
\begin{align}
    & \left.\cfxt{12}\right|_{\xi=1} = \frac{1}{2v} \left . \overline{\Gamma}^{(1)}_{ \sigma T_1T_1} \right |_{\xi=1}
    = \frac{1}{16 \pi^2 v^2} \Big [ m^2 (3 M^2+2 M_A^2)+6 (M^4+M_A^4) \Big ] \div ,  \label{theta12}\\
    & \left.
    \cfps{11}
    \right|_{\xi=1} =  \left . \overline{\Gamma}^{(1)}_{\bar c^* T_1T_1} \right |_{\xi=1} = -\frac{3 M^2+ 2 M_A^2}{8\pi^2}\div.
    \label{theta.bcT12}
\end{align}	
\end{subequations}
Using then the gauge independence of $\cfxt{12}$ we obtain, from~\1eq{T2a}
\begin{align}
	\cfct{0T_1^2}&=\frac{1}{2 m^2v} \Big ( 2 v\cfxt{12}-\overline{\Gamma}^{(1)}_{\sigma T_1T_1} \Big ) = \frac{M_A^2}{8\pi^2v^2} \div \dl,
\end{align}
which, once inserted in~\1eq{T2b} shows that 
$\cfps{11}$
is also gauge invariant, so that the condition~\noeq{T2b} reads 
\begin{align}
    -\frac{m^2}{v} \Big ( \left . \overline{\Gamma}^{(1)}_{\bar c^*T_1T_1} \right |_{\xi=0} - 
    \left . \overline{\Gamma}^{(1)}_{\bar c^*T_1T_1} \right |_{\xi=1} \Big ) =
    \left . \overline{\Gamma}^{(1)}_{\sigma T_1T_1} \right |_{\xi=0} - 
    \left . \overline{\Gamma}^{(1)}_{\sigma T_1T_1} \right |_{\xi=1}.
    \label{cc.bcTT}
\end{align}
We remark that resummation of the $T_1$-insertions is not at work for
the tadpoles in the Landau gauge
since the loop with a massless Goldstone field
in $\overline{\G}^{(1)}_{\bar c^*}$ and $\overline{\G}^{(1)}_{\sigma}$ happens to be zero in dimensional regularization.

In the Landau gauge there is no pure field redefinition
since $\left . \cfct{1} \right |_{\xi=0}=0$. On the other hand the invariant
\begin{align}
    \cfct{0}\, {\cal S}_0 \! \int \d [\sigma^*(\sigma + v) + \chi^* \chi],
\end{align}
shows that in Landau gauge also the v.e.v. $v$  renormalizes
in the same way as the field $\phi$. This is a well-known fact
in spontaneously broken gauge theories~\cite{Sperling:2013eva}.

\section{The pure external sources sector}
\label{sec:ps}

We now move to the pure external sources sector. These invariants, which are reported in~\1eq{ESinv}, cannot depend on the gauge, as we will explicitly show.

\subsection{Linear terms}

$\cfps{1}$ has been already fixed in 
Eq.(\ref{tadsb}).
$\cfps{2}$ can be fixed by looking at the $T_1$-tadpole~\noeq{GT1}:
\begin{align}
        \cfps{2} 
        = \overline{\G}^{(1)}_{T_1}  
        =  - \frac{(M^4-3 M_A^4)}{16 \pi^2}\div.
        \label{theta.T1}
\end{align}    
Notice that there are no contributions from cohomologically trivial invariants since there are no linear couplings for $T_1$ at tree-level. Consequently $\overline{\G}^{(1)}_{T_1}$ is the same both in Landau and in Feynman gauge.

\subsection{Bilinears}
   
$\cfps{3}$ is fixed by the 2-point
$\bar c^*$-amplitude~\1eq{Gc*c*}:
\begin{align}
        \cfps{3} = 
        \overline{\G}^{(1)}_{\bar c^* \bar c^*} = \frac{1}{8 \pi^2}\div  \, .
        \label{theta.bc2}
\end{align}
Notice that $\overline{\G}^{(1)}_{\bar c^* \bar c^*}$ does not develop momentum-dependent divergences and that it does not depend on the gauge. 

This is clearly not the case for $\overline{\G}^{(1)}_{T_1T_1}$ as~\1eq{GT1T1} shows; we can then read off the coefficients of the different bilinear invariants, obtaining
\begin{align}
    & \cfps{4} 
    = \overline{\G}^{(1)}_{T_1T_1}
    = \frac{3}{16 \pi^2} (M^4 + M_A^4)\div , \nonumber \\
    & \cfps{5} 
    = - \left . \frac{\partial \overline{\G}^{(1)}_{T_1T_1}}{\partial p^2} \right |_{p=0}
    = \frac{3}{32\pi^2} (M^2 + M_A^2)\div , \qquad
    & \cfps{6} = 
    \left . \frac{\partial \overline{\G}^{(1)}_{T_1T_1}}{\partial p^4} \right |_{p=0}
    = \frac{1}{32\pi^2}\div .
\end{align}

We notice that $\cfps{6}$ has been included for completeness
but does not contribute to
operators of dim. $\leq 6$ in the target theory, rather
to dim.$8$ operators.

Finally, $\cfps{7}$ has been fixed in~\1eq{theta.2bcT1}, while
the $p^2$-coefficient of the amplitude $\overline{\Gamma}^{(1)}_{\bar c^* T_1}$, see~\noeq{Gc*T1}, is 
gauge independent and implies
\begin{align}
    \cfps{8} = - 
    \left . \frac{\partial \overline{\G}^{(1)}_{T_1\bar c^*}}{\partial p^2} \right |_{p=0}=
    \frac{1}{16 \pi^2} \div .
\end{align}

\subsection{Trilinears}

While $\cfps{11}$ has been fixed in~\1eq{theta.bcT12}, it turns out that the remaining trilinears do not receive contributions from cohomologically trivial invariants. In particular we find
\begin{align}
\cfps{9} = 
 0
\end{align}
since ${\Gamma}^{(1)}_{\bar c^*\bar c^*\bar c^*}$ is UV finite, and, using the results~\noeq{Gc*c*T1} and~\noeq{GT1T1T1}
\begin{align}
    \cfps{10} &= 
    \overline{\Gamma}^{(1)}_{\bar c^* \bar c^* T_1} = -\frac{1}{4\pi^2}\div;
    &
     \cfps{12} &= 
     \overline{\G}^{(1)}_{T_1T_1T_1} = 
    -\frac{3 M^4}{4\pi^2}\div.
\end{align}

\section{The mixed external sources-field sector}\label{sec:ms}

\subsection{The $\cfxt{1}$ and $\cfxt{2}$ coefficients}

The coefficients $\cfxt{1}$ and $\cfxt{2}$ can be fixed  by evaluating the three-point functions $\overline{\G}^{(1)}_{\bar c^* \chi\chi}$ and $\overline{\G}^{(1)}_{T_1 \chi\chi}$ at zero momentum. Since
\begin{align}
& \cfct{0}\, {\cal S}_0  \int \d \left[\sigma^* (\sigma + v)  + \chi^* \chi\right]  +
    \cfct{1} {\cal S}_0  \int \d (\sigma^* \sigma  + \chi^* \chi) + 
\cfxt{1} \int \d \bar c^* \Big ( \phi^\dagger \phi - \frac{v^2}{2} \Big )
   \nonumber \\
& \supset
    \int \d  \Big ( \cfct{0} 
    + \cfct{1} +  
    \frac{\cfxt{1}}{2} 
    \Big )\bar c^* \chi^2,
\end{align}
one arrives at the relation
\begin{align}
    2 \cfct{0} 
    + 2 \cfct{1} +  
    \cfxt{1} = 
    \overline{\G}^{(1)}_{\bar c^* \chi\chi }.
\end{align}
Then, using~\3eqs{cfct.1}{cfct.0}{Gc*chch}, we immediately obtain the result
\begin{align}
    \cfxt{1} = \overline{\G}^{(1)}_{\bar c^* \chi\chi } - 2 ( \cfct{0}+\cfct{1}) = - \frac{m^2 + M^2+ M_A^2}{8 \pi^2 v^2} \div, 
    \label{theta1}
\end{align}
which, due to the compensation of the gauge parameter dependence between the amplitude and the coefficients $\cfct{0}$ and $\cfct{1}$ turns out to be gauge independent, as it should. In a similar fashion, considering the combination
\begin{align}
    & \cfct{0T_1} {\cal S}_0 \int \d T_1 \left[\sigma^* (\sigma + v)  + \chi^* \chi\right]
+ \cfxt{2} \int \d T_1 \Big ( \phi^\dagger \phi - \frac{v^2}{2} \Big ) \supset
    \int \d  
    \Big ( 
    - \cfct{0T_1} \frac{m^2}{2} +
    \frac{\cfxt{2}}{2} 
    \Big ) 
    T_1 \chi^2,
\end{align}
we get
\begin{align}
	 - \cfct{0T_1} m^2 +\cfxt{2}=\overline{\G}^{(1)}_{T_1 \chi\chi},
\end{align}
or, using the result~\noeq{GT1chch}
\begin{align}
    \cfxt{2} = 
    -\frac{m^2 (M^2 + M_A^2) + 2 (M^4 - 3 M_A^4)}{8 \pi^2 v^2}\div,
\end{align}
and again one obtains the gauge independence of this parameter as a result of the cancellation of the gauge-dependence between the 1-PI amplitude and the coefficient $\cfct{0T_1}$.

The validity of these results can be checked against the relations provided by 1-PI amplitudes involving one source and one external $\sigma$-field. For example considering the $\bar c^*\sigma$ case, we find 
\begin{align}
&   \cfct{0} {\cal S}_0  \int \d [\sigma^* (\sigma + v)  + \chi^* \chi]  + 
\cfct{1} {\cal S}_0 \Big ( \int \d (\sigma^* \sigma + \chi^* \chi) \Big ) +
\cfxt{1} \int \d \bar c^* \Big ( \phi^\dagger \phi - \frac{v^2}{2} \Big )
   \nonumber \\
& \supset
    \int \d  v \Big ( 2 \cfct{0} + \cfct{1} 
    +  \cfxt{1} 
    \Big )\bar c^* \sigma,
\end{align}
yielding the relation
\begin{align}
    v (2 \cfct{0} +  \cfct{1} + \cfxt{1}) = \overline{\G}^{(1)}_{\bar c^* \sigma},
\end{align}
which can be checked directly using~\3eqs{cfct.1}{cfct.0}{theta1}. Notice that $\overline{\G}^{(1)}_{\bar c^* \sigma}$ is the same in Feynman and Landau gauge, see~\1eq{Gc*s}; therfore, since $\cfxt{1}$ is gauge independent, so must be 
the combination $2 \cfct{0} + \cfct{1}$, as can be easily verified.

Considering the $T_1\sigma$ amplitudes, we find instead
\begin{align}
&   \cfct{0T_1} {\cal S}_0 \Big ( \int \d T_1 (\sigma^* (\sigma + v)  + \chi^* \chi) \Big ) 
+
\cfxt{2} \int \d T_1 \Big ( \phi^\dagger \phi - \frac{v^2}{2} \Big )\nonumber \\
&\supset
    \int \d  \Big ( - v m^2 \cfct{0T_1}  
    +  v \cfxt{2} 
    \Big ) T_1 \sigma.
\end{align}
Thus we get
\begin{align}
    - v m^2 \cfct{0T_1} + v \cfxt{2} = \overline{\G}^{(1)}_{T_1 \sigma},
\end{align}
which can be immediately verified using the one-loop result~\noeq{GT1s}. 

\subsection{The $\cfxt{3}$ and $\cfxt{5}$ coefficients}

In order to fix $\cfxt{3}$ and $\cfxt{5}$, we need the amplitude $\overline{\G}^{(1)}_{\bar c^* \chi \chi}$, which can be decomposed in form factors according to
\begin{align}
& \overline{\G}^{(1)}_{\bar c^* \chi \chi}(p_1,p_2) = 
\ff{0}{\bar c^* \chi\chi}+
\ff{1}{\bar c^* \chi\chi}(p_1^2+p_2^2)+
\ff{2}{\bar c^* \chi\chi}(p_1{\cdot} p_2).
\end{align}
We find
\begin{subequations}
\begin{align}
    &\cfxt{3} \int \d \bar c^* (D^\mu \phi)^\dagger D_\mu \phi \supset
    \cfxt{3} \int \d 
    \frac{\bar c^*}{2} \partial^\mu \chi \partial_\mu \chi,  \\
    & \cfxt{5} \int \d \bar c^* \Big [
    (D^2\phi)^\dagger \phi + \mathrm{h.c.}\Big ] \supset 
    \cfxt{5} \int \d
    \bar c^* \chi \square \chi,
\end{align}
\end{subequations}
which, using the result~\1eq{Gc*chch}, implies the following identifications
\begin{align}
    \cfxt{3} &= -\ff{2}{\bar c^* \chi\chi} =
    - \frac{1}{16 \pi^2}
    \frac{g}{v \Lambda}
    \Big ( 2 + \frac{g v}{\Lambda} \Big )
    \div;&
    \cfxt{5} &= -\ff{1}{\bar c^* \chi\chi} =
    - \frac{1}{16 \pi^2}
    \frac{g}{\Lambda v}\div 
    \, . 
\end{align}
Notice that both coefficients are the same in Landau and Feynman gauge, as expected.

In this case a consistency check is provided by the three-point function $\overline{\Gamma}^{(1)}_{\bar c^* A_\mu A_\nu}$, since one has
\begin{align}
& \cfxt{3} \int \d \bar c^* (D^\mu \phi)^\dagger D_\mu \phi + \cfxt{5} \int \d \bar c^* \Big [
    (D^2\phi)^\dagger \phi + h.c.\Big ] \supset 
     \int \d
    \frac{M_A^2}{2}  \Big ( \cfxt{3} - 2 \cfxt{5} \Big ) \bar c^* A^2,
\end{align}
so that
\begin{align}
    M_A^2 ( \cfxt{3} - 2 \cfxt{5} ) g_{\mu\nu} =
    \left . 
    \overline{\G}^{(1)}_{\bar c^* A_\mu A_\nu}(p_1,p_2)
    \right |_{p_1=p_2=0},
\end{align}
as can be easily verified with the help of~\1eq{Gc*AA}.

\subsection{The $\cfxt{4}$ and $\cfxt{6}$ coefficients}

In order to fix $\cfxt{4}$ and $\cfxt{6}$ we need the amplitude $\overline{\G}^{(1)}_{T_1 \chi \chi}$, which we decompose as before according to
\begin{align}
\overline{\G}^{(1)}_{T_1 \chi \chi} (p_1,p_2)= 
\ff{0}{T_1 \chi\chi}+
\ff{1}{T_1 \chi\chi}(p_1^2+p_2^2)+
\ff{2}{T_1 \chi\chi}({p_1{\cdot}p_2}) 
+ {\cal O}(p_i^4),
\end{align}
and the dots denote terms of order $p^4$, which are not needed.

There are two projections to be considered, namely $T_1\partial^\mu\chi\partial_\mu \chi$ and  $T_1 \chi \square \chi$, to which the cohomologically trivial invariants can also contribute. To begin with, observe that 
%
\begin{align}
	\cfct{1} {\cal S}_0 \int \d \frac{1}{1+T_1}(\sigma^* \sigma + \chi^* \chi) 
 	& = \cfct{1} {\cal S}_0\int \d 
 	(1 - T_1 + \cdots)  (\sigma^* \sigma +  \chi^* \chi) \nonumber \\
 	& \supset\cfct{1} \int \d \Big (T_1 \partial^\mu\chi \partial_\mu\chi 
 	+ T_1 \chi \square \chi 
 	\Big ).
\end{align}
On the other hand we have
\begin{align}
    & \cfct{0} {\cal S}_0
        \int \d 
    ( \sigma^* (\sigma + v) + \chi^* \chi )
    +
    \cfct{0T_1} {\cal S}_0
    \int \d T_1
    [ \sigma^* (\sigma + v) + \chi^* \chi ]  \nonumber \\
    & \supset
    \int \d \Big [ 
     \cfct{0} T_1 
    \partial^\mu \chi \partial_\mu \chi
    - \cfct{0T_1} T_1 \chi \square \chi
 \Big ].
\end{align}

Therefore we obtain
\begin{align}
\cfct{1}-\cfct{0T_1}+\cfxt{4} &=  -\ff{2}{T_1 \chi\chi};& 
2 (\cfct{1} + \cfct{0} ) +\cfxt{6} = - \ff{1}{T_1 \chi\chi} \, ,
\end{align}
from which, using~\1eq{GT1chch}, we finally get the values
\begin{subequations}
\begin{align}
   & \cfxt{4} = 
    -\frac{1}{32 \pi^2 v^2}
    \Big [ 
    4 m^2 + M_A^2  \Big ( 4 - 3 \frac{g^2 v^2}{\Lambda^2} \Big ) +
    M^2  \Big ( 4 + 3 \frac{g^2 v^2}{\Lambda^2} \Big )
    \Big ]\div, \nonumber \\
    & \cfxt{6} = -
    \frac{1}{16 \pi^2 v^2}
    \Big [ m^2 - M_A^2+ M^2 \Big (1 + 2 \frac{gv}{\Lambda} \Big ) \Big ]
    \div.
\end{align}
\end{subequations}

Similarly to what we have done in the previous case, we can check the results above using the three-point function $\overline{\Gamma}^{(1)}_{T_1 A_\mu A_\nu}$. Indeed we have
\begin{align}
& \cfxt{4} \int \d T_1 (D^\mu \phi)^\dagger D_\mu \phi + \cfxt{6} \int \d T_1 \Big [
    (D^2\phi)^\dagger \phi + \mathrm{h.c.}\Big ]  + \cfct{0} {\cal S}_0 \int \d [\sigma^* (\sigma + v)  + \chi^* \chi] \nonumber \\
    &+\cfct{0T_1} {\cal S}_0 \int \d T_1 [\sigma^* (\sigma + v)  + \chi^* \chi]
    \supset \int \d
    \frac{M_A^2}{2} [\cfxt{4} - 2 \cfxt{6} +
    2 ( \cfct{0} + \cfct{0T_1} )] T_1 A^2,
\end{align}
implying the consistency condition
\begin{align}
    M_A^2 \Big [  \cfxt{4} - 2 \cfxt{6} +
    2 ( \cfct{0} + \cfct{0T_1} )
    \Big ] g_{\mu\nu} =
    \left . 
    \overline{\G}^{(1)}_{T_1 A_\mu A_\nu}(p_1,p_2)
    \right |_{p_1=p_2=0}.
\end{align}
the validity of which can be easily verified with the help of~\1eq{GT1AA}.

\subsection{The $\cfxt{7}$ and $\cfxt{8}$ coefficients}

In this sector the relevant projections are
\begin{subequations}
\begin{align}
	&   \cfct{0} {\cal S}_0 \int \d [\sigma^* (\sigma + v)  + \chi^* \chi] +
	\cfct{1} {\cal S}_0 \int \d (\sigma^* \sigma + \chi^* \chi)+
	\cfxt{1} \int \d \bar c^* \Big ( \phi^\dagger \phi - \frac{v^2}{2} \Big ) \nonumber \\
	&+ \cfxt{7}\int \d  \bar c^*
 	\Big ( \phi^\dagger \phi - \frac{v^2}{2} \Big )^2\supset
    \int \d   
    \Big ( \cfct{0} + \cfct{1} + \frac{\cfxt{1}}{2}
    + v^2 \cfxt{7}
     \Big ) \bar c^* \sigma^2,\\
	& \cfct{0T_1}{\cal S}_0\int \d  T_1 [\sigma^* (\sigma+v) + \chi^* \chi]
	- \cfct{1} {\cal S}_0 \int \d 
	 T_1 (\sigma^* \sigma + \chi^* \chi) 
	+\cfxt{2} \int \d T_1\Big ( \phi^\dagger \phi - \frac{v^2}{2} \Big ) \nonumber \\
	& + \cfxt{8}\int \d   T_1 \Big ( \phi^\dagger \phi - \frac{v^2}{2} \Big )^2\supset
	\int \d \Big ( m^2 \cfct{1}- \frac{5}{2} \cfct{0T_1} m^2+ \frac{\cfxt{2}}{2} +v^2 \cfxt{8}\Big ) T_1 \sigma^2,
\end{align}	
\end{subequations}
yielding the relations
\begin{align}
    & 2(\cfct{0} + \cfct{1}) + \cfxt{1}
    +  2 v^2 \cfxt{7} =
    \overline{\G}^{(1)}_{\bar c^* \sigma\sigma};&
    & 2 m^2 \cfct{1}- 5 \cfct{0T_1} m^2+\cfxt{2} +2 v^2 \cfxt{8} = \overline{\G}^{(1)}_{T_1 \sigma \sigma} \, ,
\end{align}
and, finally, the values
\begin{align}
    &\cfxt{7} = 0;&
    &\cfxt{8} = -\frac{1}{8\pi^2 v^4}
    \Big [ m^4+2m^2(M^2+M_A^2) + 2 (M^4- 3 M_A^4) \Big ] \div,
\end{align}
see~\2eqs{Gc*ss}{GT1ss}.

\subsection{The $\cfxt{9}$ and $\cfxt{10}$ coefficients}

The fact that the function $\overline{\G}^{(1)}_{\bar c^* A_\mu A_\nu}$ turns out to be momentum independent, see~\1eq{Gc*AA}, implies immediately that
\begin{align}
    \cfxt{9}=0.
\end{align}
Next, in order to extract the coefficient $\cfxt{10}$ one needs first to change the variables to the contractible pairs basis, as explained in~\cite{BQ:2019a}. To this end, one replaces the derivatives of the gauge field with a linear combination of the complete symmetrization over the Lorentz indices and a contribution depending on the field strength:
\begin{align}
    \partial_{\nu_1 \dots \nu_\ell} A_\mu = \partial_{(\nu_1 \dots \nu_\ell} A_{\mu)}+\frac{\ell}{\ell+1}\partial_{(\nu_1 \dots \nu_{\ell-1}} F_{\nu_\ell) \mu},
    \label{cp.gauge}
\end{align}
where $(\dots)$ denote complete symmetrization. In the present case it is therefore sufficient to consider the monomial $T_1 \partial^\mu A^\nu \partial_\mu A_\nu$ since, due to~\1eq{cp.gauge} 
\begin{align}
	T_1 \partial^\mu A^\nu \partial_\mu A_\nu = 
	T_1 \partial^{(\mu} A^{\nu)}\partial_{(\mu} A_{\nu)}  +
	\frac{T_1}{4}  F^{\mu\nu} F_{\mu\nu}. 
\end{align}

Then, after we decompose the amplitude $\overline{\G}^{(1)}_{T_1 A_\mu A_\nu}$
according to
\begin{align}
    \overline{\G}^{(1)}_{T_1 A_\mu A_\nu}(p_1,p_2) & =  [ \gamma^0_{T_1 AA}  -2 \gamma^1_{T_1 AA} (p_1{\cdot} p_2) +
    \gamma^2_{T_1 AA} (p_1^2+p_2^2)] g^{\mu\nu} \nonumber \\
    & + \gamma^3_{T_1 AA}
    p_1^\mu p_2^\nu + 
    \gamma^4_{T_1 AA}
    p_1^\nu p_2^\mu,
\end{align}
\1eq{GT1AA} gives
\begin{align}
    \cfxt{10}= \frac{\gamma^1_{TAA}}{4} =
    -\frac{M_A^2}{128 \pi^2}\frac{g^2}{v^2 \Lambda^2} \div.
    \label{cfxt.10}
\end{align}

\subsection{The $\cfxt{11}$, $\cfxt{12}$ and $\cfxt{13}$ coefficients}

The coefficient $\cfxt{12}$ has been fixed already, see~\1eq{theta12}; on the other hand, $\cfxt{11}$ is determined by the projection of
\begin{align}
&   \cfxt{11} \int \d
    \bar c^* T_1 
    \Big ( \phi^\dagger \phi - \frac{v^2}{2} \Big ) 
    - \cfct{1} {\cal S}_0\int \d T_1 (\sigma^* \sigma + \chi^* \chi)+
    \cfct{0T_1}
    {\cal S}_0   \int \d T_1 [\sigma^* (\sigma+v) + \chi^* \chi]
\nonumber \\
& \supset \int \d
\Big ( v \cfxt{11} - v \cfct{1} + 2 v \cfct{0T_1} \Big ) \bar c^* T_1 \sigma \, .
\end{align}
yielding
\begin{align}
	\cfxt{11} = \frac{1}{v}
	\Big ( \overline{\G}^{(1)}_{\bar c^* T_1 \sigma} + v \cfct{1} - 2 v \cfct{0T_1} \Big ) =
	\frac{1}{4 \pi^2 v^2}(m^2+ M^2 + M_A^2)\div,
\end{align}
where the one-loop result~\noeq{Gc*T1s} has been used.
Finally,
\begin{align}
    \cfxt{13} \int \d 
    (\bar c^*)^2 \Big (
    \phi^\dagger \phi - \frac{v^2}{2}
    \Big ) \supset \int \d
    \cfxt{13} v  \sigma (\bar c^*)^2,
\end{align}
which implies\begin{align}
    \cfxt{13}= \frac{1}{2v} 
    \overline{\G}^{(1)}_{\bar c^*\bar c^* \sigma } = 0,
\end{align}
as this amplitude turns out to be UV finite.

\section{The gauge-invariant field sector}
\label{sec:gi}

The last sector we need to consider is finally the one of gauge invariants containing only the fields.

\subsection{The $\cfgi{2}$ and $\cfgi{3}$ coefficients}
While the coefficient $\cfgi{1}$ has been already fixed, see~\1eq{l1.bc},
$\cfgi{2}$ and $\cfgi{3}$ can be determined by considering the two- and three-point $\sigma$ amplitudes. The relevant projection equation are
\begin{subequations}
\begin{align}
	&   \cfct{0} {\cal S}_0 \int \d [\sigma^* (\sigma + v)  + \chi^* \chi] +
	\cfct{1} {\cal S}_0 \int \d (\sigma^* \sigma + \chi^* \chi)+
	\cfgi{1} \int \d 
	\Big ( \phi^\dagger \phi - \frac{v^2}{2} \Big ) \nonumber \\
	&+\cfgi{2} \int \d 
	\Big ( \phi^\dagger \phi - \frac{v^2}{2} \Big )^2\supset \int \d \Big( v^2\cfgi{2}+\frac12\cfgi{1}- m^2 \cfct{1} - \frac52 m^2\cfct{0}\Big)\sigma^2, \\
	&   \cfct{0} {\cal S}_0 \int \d [\sigma^* (\sigma + v)  + \chi^* \chi] +
	\cfct{1} {\cal S}_0 \int \d (\sigma^* \sigma + \chi^* \chi)+
	\cfgi{2} \int \d 
	\Big ( \phi^\dagger \phi - \frac{v^2}{2} \Big )^2 \nonumber \\
	&+\cfgi{3} \int \d 
	\Big ( \phi^\dagger \phi - \frac{v^2}{2} \Big )^3\supset \int \d \Big(2 v^3 \cfgi{3} + 2 v \cfgi{2}
- \frac{3 m^2}{v} \cfct{1} -  \frac{4 m^2}{v} \cfct{0}\Big )\sigma^3,
\end{align}
\end{subequations}
yielding
\begin{subequations}
\begin{align}
& 2 v^2\cfgi{2}+\cfgi{1}-2 m^2 \cfct{1} - 5 m^2\cfct{0} =
\overline{\G}^{(1)}_{\sigma\sigma},\\
& 6 v^3 \cfgi{3} + 6 v \cfgi{2}
- \frac{9 m^2}{v} \cfct{1} -  \frac{12 m^2}{v} \cfct{0} = 
\overline{\G}^{(1)}_{\sigma\sigma\sigma}.
\end{align}
\end{subequations}

\2eqs{Gss}{Gsss} implies then the following results
\begin{subequations}
	\begin{align}
	\cfgi{2} & = 	\frac{1}{16 \pi^2 v^4}
    \Big [ m^4 + 2 m^2 (M^2 + M_A^2) + 2 (M^4+3 M_A^4) \Big ] \div, \nonumber \\
	\cfgi{3} & =	 0.
\end{align}\end{subequations}
The values of these coefficients  can be checked by looking at the  $\overline{\G}^{(1)}_{\sigma \chi\chi}$
and $\overline{\G}^{(1)}_{\sigma\sigma \chi\chi}$ amplitudes, for which the projection equation
\begin{align}
	&   \cfct{0} {\cal S}_0 \int \d [\sigma^* (\sigma + v)  + \chi^* \chi] +
	\cfct{1} {\cal S}_0 \int \d (\sigma^* \sigma + \chi^* \chi)+\cfgi{2} \int \d 
	\Big ( \phi^\dagger \phi - \frac{v^2}{2} \Big )^2 \nonumber \\
	&\supset \int\d \Big( v \cfgi{2} - \frac{3 m^2}{2v} \cfct{1} - \frac{2 m^2}{v} \cfct{0} \Big)\sigma\chi^2+\int\d \Big(\frac12 \cfgi{2} - \frac{ m^2}{v^2} \cfct{1}- \frac{ m^2}{v^2} \cfct{0}\Big)\sigma^2\chi^2,
\end{align}
gives rise to the consistency conditions
\begin{align}
& 2 v \cfgi{2} - \frac{3 m^2}{v} \cfct{1} - \frac{4 m^2}{v} \cfct{0} = \overline{\G}^{(1)}_{\sigma \chi\chi},
\nonumber \\
& 2 \cfgi{2} - \frac{4 m^2}{v^2} \cfct{1}
- \frac{4 m^2}{v^2} \cfct{0}
=
\overline{\G}^{(1)}_{\sigma\sigma \chi\chi},
\end{align}
which, using~\2eqs{Gschch}{Gsschch}, can be easily proven to be fulfilled.

\subsection{The $\cfgi{4}$ and $\cfgi{5}$ coefficients}

These coefficients are fixed by the 2-point Goldstone amplitude, which is controlled by the invariants
\begin{align}
    & \cfct{0} {\cal S}_0 \int \d [\sigma^* (\sigma + v)  + \chi^* \chi] +
	\cfct{1} {\cal S}_0 \int \d (\sigma^* \sigma + \chi^* \chi)  +
    \cfgi{1} \int \d \Big ( \phi^\dagger \phi - \frac{v^2}{2} \Big ) 
    \nonumber \\
    &+ \cfgi{4} \int \d (D^\mu \phi)^\dagger D_\mu \phi
     +\cfgi{5} \int \d \phi^\dagger
  [ (D^2)^2 + 
  D^\mu D^\nu D_\mu D_\nu + 
  D^\mu D^2 D_\mu ] \phi 
    \nonumber \\
    & \supset
    \int \d 
    \Big [
    \frac{1}{2} \Big ( \cfgi{1} - m^2 \cfct{0} \Big ) \chi^2 +
    \Big ( \cfct{0}+\cfct{1}+\frac{\cfgi{4}}{2} \Big )\partial^\mu \chi \partial_\mu \chi
    + \frac{3}{2} \cfgi{5} \chi \square^2 \chi 
    \Big ],
\end{align}
which gives rise to the following projections
\begin{align}
& \cfgi{1} -  m^2 \cfct{0} = \left . \overline{\G}^{(1)}_{\chi\chi} \right |_{p^2=0};&
& 2(\cfct{0}+\cfct{1})+\cfgi{4}
= \left . \frac{\partial\overline{\G}^{(1)}_{\chi\chi}}{\partial p^2} \right |_{p^2=0};&
&3 \cfgi{5} = \left . \frac{\partial\overline{\G}^{(1)}_{\chi\chi}}{\partial (p^2)^2} \right |_{p^2=0}.
\end{align}
From the one-loop expression reported in~\noeq{Gchch}, we then obtain the gauge-independent coefficients
\begin{subequations}
\begin{align}
 &  \cfgi{4} = -
 \frac{1}{32 \pi^2 v^2}
 \Big [ \frac{gv}{\Lambda} \Big (4 - \frac{gv}{\Lambda} \Big )M^2
 + M_A^2 \Big (16  + 
 12 \frac{gv}{\Lambda} + 3 \frac{g^2 v^2}{\Lambda^2}
 \Big ) \Big ] \div,
 \label{l4} \\
 & \cfgi{5} =  \frac{g^2}{96 \pi^2 \Lambda^2} \div .\label{l5}
\end{align}	
\end{subequations}

\subsection{The $\cfgi{6}$ and $\cfgi{7}$ coefficients}

The relevant Green's function for fixing these coefficients is the four-point
Goldstone amplitude since
\begin{align}
&\cfct{0} {\cal S}_0 \int \d [\sigma^* (\sigma + v)  + \chi^* \chi] +
	\cfct{1} {\cal S}_0 \int \d (\sigma^* \sigma + \chi^* \chi)+\cfgi{2} \int \d 
	\Big ( \phi^\dagger \phi - \frac{v^2}{2} \Big )^2
\nonumber \\
&
+\cfgi{6} \int \d \Big ( \phi^\dagger \phi- \frac{v^2}{2} \Big ) \Big ( \phi^\dagger D^2 \phi + (D^2\phi)^\dagger \phi \Big ) +\cfgi{7} \int \d \Big ( \phi^\dagger \phi- \frac{v^2}{2} \Big ) (D^\mu \phi)^\dagger D_\mu \phi 
   \nonumber \\
& \supset  \int \d 
  \Big \{ \Big [ \frac{\cfgi{2}}{4}   - (\cfct{0}+\cfct{1}) \frac{m^2}{2v^2}
  \Big ]  \chi^4 +  \frac{\cfgi{6}}{2} \chi^3 \square \chi
  + \frac{\cfgi{7}}{4} \chi^2 \partial^\mu \chi \partial_\mu \chi
  \Big \},
\end{align}
yielding
\begin{align}
    6 \cfgi{2} -  \frac{12 m^2}{v^2} (\cfct{0}+ \cfct{1}) 
    - 3 \cfgi{6} \sum_{i=1}^4
    p_i^2
    - \cfgi{7} 
    \sum_{i<j} p_i p_j
= 
\overline{\G}^{(1)}_{\chi\chi\chi\chi}(p_i).
\label{4chi.proj}
\end{align}
Notice that we keep the momentum dependence of the four point $\chi$ amplitude on the right-hand side. A remark is in order here. Before attempting to extract the coefficients of the  momenta polynomia on the left-hand side of~\1eq{4chi.proj}, we need to take into account the fact that {\tt FeynArts} and {\tt FormCalc} internally implement momentum conservation, so the amplitude is only known on the hyperplane $\sum_i p_i = 0$. Hence we eliminate $p_4$ in favor of the remaining momenta, $p_4 = -\sum_{i=1}^3 p_i$, so that~\1eq{4chi.proj} becomes
\begin{align}
 6 \cfgi{2} -  \frac{12 m^2}{v^2} (\cfct{0}+ \cfct{1}) 
  -
  ( 6 \cfgi{6} - \cfgi{7} )
  \Big ( \sum_{i=1}^3 p_i^2 + \sum_{i<j} p_i p_j \Big )
= 
\overline{\G}^{(1)}_{\chi\chi\chi\chi}(p_1,p_2,p_3).
\label{4chi.mom.cons}
\end{align}
Then the condition
\begin{align}
       6 \cfgi{2} -  \frac{12 m^2}{v^2} (\cfct{0}+ \cfct{1}) =
       \left . \overline{\G}^{(1)}_{\chi\chi\chi\chi}  \right |_{p_i=0},
\end{align}
is easily verified, see~\1eq{Gchchchch}. On the other hand, we notice that the restriction of $\overline{\G}^{(1)}_{\chi^4}$ on the momentum conservation hyperplane only fixes the combination $6 \cfgi{6} - \cfgi{7} $, and an additional amplitude needs to be considered to fix the two coefficients separately.

To this end, let us consider the two point $\sigma$-amplitude, with the following projection on the derivative-dependent sector
\begin{align}
	& \cfct{0} {\cal S}_0 \int \d [\sigma^* (\sigma + v)  + \chi^* \chi] +
	\cfct{1} {\cal S}_0 \int \d (\sigma^* \sigma + \chi^* \chi)  +
     \cfgi{4} \int \d (D^\mu \phi)^\dagger D_\mu \phi\nonumber \\
    &+\cfgi{6} \int \d \Big ( \phi^\dagger \phi- \frac{v^2}{2} \Big ) \Big ( \phi^\dagger D^2 \phi + (D^2\phi)^\dagger \phi \Big )\supset\int \d \Big [ \Big ( \frac{\cfgi{4}}{2} + \cfct{0} + \cfct{1} \Big )\partial^\mu \sigma \partial_\mu \sigma
   + v^2 \cfgi{6} \sigma \square \sigma \Big ],
\end{align}
leading to the condition
\begin{align}
   2 v^2 \cfgi{6} - \cfgi{4} 
-2 ( \cfct{0} + \cfct{1} ) = -
\left . \frac{\partial \overline{\G}_{\sigma\sigma}^{(1)}}{\partial p^2} \right |_{p^2=0}.
\end{align}
This gives then the result, see~\1eq{Gss}
\begin{align}
    \cfgi{6} = \frac{1}{64 \pi^2 v^3}
    \frac{g}{\Lambda}
    \Big [ 4 m^2 + (M^2 - 3 M_A^2)
    \Big ( 4 + \frac{gv}{\Lambda} \Big )
    \Big ] \div,
\end{align}
which in combination with~\2eqs{4chi.mom.cons}{Gchchchch} yields finally
\begin{align}
    \cfgi{7}=
    \frac{1}{32 \pi^2 v^3}
    \frac{g}{\Lambda}
    \Big [ 2 m^2 \Big ( 2 + \frac{gv}{\Lambda} \Big )
    - M^2 \Big ( 4 - 5 \frac{gv}{\Lambda} \Big ) - 3 M_A^2 \Big ( 
    12 + 5 \frac{gv}{\Lambda}
    \Big )
    \Big ] \div.
\end{align}
We can check this result against the projections on the monomials $\sigma \chi \square \chi, \sigma \partial^\mu \chi \partial_\mu \chi$, namely  (we use integration by parts in the last line)
\begin{align}
& \cfgi{6} \int \d  \Big ( \phi^\dagger \phi -
\frac{v^2}{2} \Big ) 
\Big ( \phi^\dagger D^2 \phi +
(D^2\phi)^\dagger \phi
\Big ) + 
\cfgi{7} \int \d  \, \Big ( \phi^\dagger \phi -
\frac{v^2}{2} \Big ) (D^\mu \phi)^\dagger D_\mu \phi 
 \nonumber \\
& \supset\int \d \, \Big ( v \cfgi{6}
\sigma \chi \square \chi + 
\frac{v \cfgi{7}}{2} \sigma \partial^\mu \chi \partial_\mu \chi
+ \frac{\cfgi{6} v}{2} \chi^2 \square \sigma 
\Big ) \nonumber \\
& =
 \int \d \, \Big [
2 v \cfgi{6} \sigma \chi \square \chi
+
\Big ( 
 v \cfgi{6} + \frac{v \cfgi{7}}{2} 
\Big )
 \sigma \partial^\mu \chi \partial_\mu \chi
\Big ].
\label{proj.sigma2chi}
\end{align}
After eliminating the $\sigma$-momentum in favour of the remaining tow by using momentum conservation, the resulting amplitude can be expanded as
\begin{align}
\overline{\G}^{(1)}_{\sigma \chi \chi}(p_1,p_2) =
 \gamma_{\sigma\chi\chi}+
 \gamma^1_{\sigma\chi\chi} (p_1^2 + p_2^2)
 +  \gamma^2_{\sigma\chi\chi} p_1{\cdot} p_2 + {\cal O}(p^4)
\end{align}
\1eq{proj.sigma2chi} then implies the consistency conditions
\begin{align}
    &  2 v\cfgi{6}+ \gamma^1_{\sigma\chi\chi} = 0;&
     2 v \cfgi{6}  + v \cfgi{7}
   + \gamma^2_{\sigma\chi\chi} = 0,
\end{align}
which can be easily verified using the result~\1eq{Gschch}.

\subsection{The $\cfgi{8}$ and $\cfgi{9}$ coefficients}

These coefficients are controlled by the $AA$ amplitude which also provides a non-trivial example of the contractible pairs technique. Indeed, the two-point function of the Goldstone field fixes the coefficient $\cfgi{5}$ via the projection on the monomial $\int \d \chi \square^2 \chi$; on the other hand, the $\cfgi{5}$ invariant admits also a non-trivial expansion in power of the gauge field, precisely accounting for the non-transverse form factors of $\overline{\G}^{(1)}_{A^\mu A^\nu}$.

To see this in detail, observe that the relevant invariants are
\begin{align}
	& \cfct{0} {\cal S}_0 \int \d [\sigma^* (\sigma + v)  + \chi^* \chi] +
	\cfct{1} {\cal S}_0 \int \d (\sigma^* \sigma + \chi^* \chi) \nonumber \\
	&+\cfgi{5} \int \d \phi^\dagger[ (D^2)^2 + D^\mu D^\nu D_\mu D_\nu +  D^\mu D^2 D_\mu ] \phi + \frac{\cfgi{8}}{2} \int \d F_{\mu\nu}^2 +
  \cfgi{9}  \int \d \partial^\mu F_{\mu\nu} \partial_\rho F^{\rho\nu}\nonumber \\
  & \supset
  \int \d \Big [
  \Big ( \cfct{0} +  \frac{\cfgi{4}}{2}  \Big ) e^2 v^2 A^2 
  - \frac{\cfgi{5}}{2} e^2 v^2 
  (2 A^\mu \partial_\mu \partial A + 
  A^\mu \square A_\mu ) + 
  \frac{\cfgi{8}}{2} (\partial^\mu A^\nu - \partial^\nu A^\mu)^2\nonumber \\
  & 
  + \cfgi{9} (\square A^\mu - \partial^\mu (\partial A))^2
  \Big ]
\end{align}
There are no contribution of order $p^4$ in $\overline{\G}^{(1)}_{A^\mu A^\mu}$, see~\1eq{GAA}, so
\begin{align}
    \cfgi{9}=0.
\end{align}
The remaining terms give the projection equation
\begin{align}
     \Big [   e^2 v^2 ( 2 \cfct{0} +  \cfgi{4}) + ( 2 \cfgi{8} + e^2 v^2
     \cfgi{5}) p^2  \Big ] g^{\mu\nu} 
     + 2 \Big ( e^2 v^2 \cfgi{5} - \cfgi{8}  \Big ) p^\mu p^\nu 
     = \overline{\G}^{(1)}_{A^\mu A^\nu}(p).
\end{align}
Notice that in the right-hand side of the above equation we keep the momentum dependence of the two point gauge amplitude. From~\4eqs{cfct.0}{l4}{l5}{GAA}, we see that the above equation is verified with
\begin{align}
\cfgi{8} = - \frac{M_A^2}{96 \pi^2 v^2}
\Big ( 2 + 2 \frac{gv}{\Lambda} + 
 \frac{g^2 v^2}{\Lambda^2} \Big ) \div,
\end{align}
which implies that $\cfgi{8}$ is gauge-independent, as it should.

\subsection{The $\cfgi{10}$ coefficient}

This coefficient can be obtained in much the same way as $\cfxt{10}$, {\it i.e.}, by the contractible pair method. Parameterize the amplitude $\overline{\G}^{(1)}_{\sigma A_\mu A_\nu}$ according to
\begin{align}
    \overline{\G}^{(1)}_{\sigma A_\mu A_\nu}(p_1,p_2)= & [ \gamma^0_{\sigma AA}  -2 \gamma^1_{\sigma AA} p_1\cdot p_2 +
    \gamma^2_{\sigma AA} (p_1^2+p_2^2)] g^{\mu\nu}
    + \gamma^3_{\sigma AA} 
    p_1^\mu p_2^\nu + 
    \gamma^4_{\sigma AA} 
    p_1^\nu p_2^\mu, 
\end{align}
and extract $\cfgi{10}$ through the form factor $\gamma^1_{\sigma AA}$:
\begin{align}
    \cfgi{10} =  
    \int \d F^2_{\mu\nu}
    \Big ( \phi^\dagger \phi - \frac{v^2}{2} \Big ) \supset\cfgi{10}
    \int \d 2 \sigma \partial^\mu A^\nu \partial_\mu A_\nu.       
\end{align}
We obtain, see~\1eq{GsAA},
\begin{align}
    \cfgi{10}= \frac{\gamma^1_{\sigma AA}}{4v} =
    \frac{M_A^2}{128 \pi^2}\frac{g^2}{v^2\Lambda^2}
    \Big ( -4 + \frac{gv}{\Lambda} \Big )
    \div.
\end{align}
Notice in particular that the combination 
\begin{align}
    \cfgi{10}+\frac{g}{v\Lambda} \cfxt{10} = - \frac{M_A^2}{32 \pi^2} \frac{g^2}{\Lambda^2 v^2}\div,
\end{align}
correctly reproduces the coefficient $c_{\cal O}^{(1)}$ of~\cite{BQ:2019a}.

\section{Mapping}

\subsection{\label{sec:map}Renormalization coefficients}

We are now in a position to evaluate the renormalization coefficients of the operators of dimension less or equal to $6$ in the target theory. For that purpose one simply needs to map the invariants depending on the external sources by applying the substitution rules~\noeq{repl.fin.1} and  collecting the contributions to the operator one is interested in. 

Notice that all the coefficients obtained must be gauge-invariant (as a consequence of the gauge-invariance of the $\cfxt{i},\cfps{i}$ and $\cfgi{i}$ coefficients); in addition they must not depend on $m^2$. The latter is a highly non-trivial check of the computations, due to the ubiquitous presence of $m^2$ in the projections as well as in the amplitudes.

In what follows, we list here the results for all possible operators, reinstating the correct $D$-dimensional dependence on the 't Hooft mass $\mu$.
\newpage
\begin{itemize}
\item $\phi^\dagger \phi - \frac{v^2}{2}$
    
\begin{align}
   \widetilde{\lambda}_1&=\frac{1}{v^2} \Big [(M^2-m^2) \cfps{1} + \frac{gv}{\Lambda} \cfps{2} + v^2 \cfgi{1}\Big ] & \nonumber \\
   &=\frac{\mu^{-\epsilon}}{16 \pi^2 v^2} \Big \{ M^4 \Big ( 3 - \frac{gv}{\Lambda} \Big ) + M_A^2 \Big [ M^2 + 3 M_A^2 \Big (2 + \frac{gv}{\Lambda} \Big ) \Big ]\Big \} \div.
\end{align}

\item $ \Big ( \phi^\dagger \phi - \frac{v^2}{2} \Big )^2$

\begin{align}
  \widetilde{\lambda}_2&= \frac{(m^2-M^2)^2}{2 v^4}
  \cfps{3} +
  \frac{g^2}{2\Lambda^2 v^2}
  \cfps{4} +
  \frac{g}{\Lambda v^3}(m^2- M^2)\cfps{7} + 
  \frac{m^2-M^2}{v^2}
  \cfxt{1}+ \frac{g}{\Lambda v} \cfxt{2} + \cfgi{2}
  \nonumber \\
  &= \frac{\mu^{-\epsilon}}{32 \pi^2 v^4}
  \Big \{ 
  4 M_A^2 M^2 \Big ( 1 - \frac{gv}{\Lambda} \Big )
  + 3 M_A^4 \Big ( 
  4 + 8 \frac{gv}{\Lambda} +
  \frac{g^2 v^2}{\Lambda^2}
  \Big )\nonumber \\
  &+
  M^4 \Big ( 10 -12 \frac{gv}{\Lambda} + 3 \frac{g^2 v^2}{\Lambda^2} \Big ) 
  \Big \} \div.
\end{align}
\item $\Big ( \phi^\dagger \phi - \frac{v^2}{2} \Big )^3$
\begin{align}
    \widetilde{\lambda}_3&= \frac{(m^2-M^2)^3}{6 v^6} \cfps{9} +
    \frac{g (m^2-M^2)^2}{2 \Lambda v^5}\cfps{10}
    + \frac{g^2(m^2- M^2)}{2 \Lambda^2 v^4}\cfps{11}+
    \frac{g^3}{6 \Lambda^3 v^3}
    \cfps{12} \nonumber \\ 
    & +
    \frac{m^2-M^2}{v^2}\cfxt{7} +
    \frac{g}{v\Lambda}\cfxt{8}+
    \frac{g(m^2-M^2)}{\Lambda v^3}\cfxt{11}+
    \frac{g^2}{\Lambda^2 v^2}\cfxt{12} +
    \frac{(m^2-M^2)^2}{v^4}
    \cfxt{13} + \cfgi{3} \nonumber \\
    & = - \frac{\mu^{-\epsilon}}{16 \pi^2 v^5}
    \frac{g}{\Lambda} 
    \Big [ 
    2 M^2 M_A^2 \Big( 2 - \frac{gv}{\Lambda} \Big )
    - 6 M_A^4\Big( 2 + \frac{gv}{\Lambda} \Big )
    + M^4
    \Big( 10 -9 \frac{gv}{\Lambda}
    + 2 \frac{g^2v^2}{\Lambda^2}\Big )
    \Big ] \div.
\end{align}
\item
$(D^\mu \phi)^\dagger D_\mu \phi$
\begin{align}
\widetilde\lambda_4&=\frac{g}{\Lambda v}\cfps{1} + \cfgi{4} \nonumber \\
&=-\frac{\mu^{-\epsilon}}{32 \pi^2 v^2}
\Big [ M^2  \frac{gv}{\Lambda} 
\Big ( 6 - \frac{gv}{\Lambda} \Big ) +
M_A^2
\Big (
16 + 14 \frac{gv}{\Lambda}
+
3\frac{g^2v^2}{\Lambda^2}
\Big ) 
\Big ]\div.
\end{align}
\item 
$\phi^\dagger  [ (D^2)^2 + D^\mu D^2 D_\mu + D^\mu D^\nu D_\mu D_\nu  ] \phi$
\begin{align}
   \widetilde{\lambda}_5&\equiv \cfgi{5} = \frac{\mu^{-\epsilon}}{96 \pi^2}\frac{g^2}{\Lambda^2} \div .
\end{align}
\item $\Big ( \phi^\dagger \phi - \frac{v^2}{2} \Big ) ( \phi^\dagger D^2\phi + \mathrm{h.c.})$
\begin{align}
\widetilde{\lambda}_6&=   \frac{g^2}{2\Lambda^2v^2}
    \cfps{5} +
    \frac{g}{\Lambda v^3}(m^2- M^2)
    \cfps{8}+
    \frac{m^2-M^2}{v^2} \cfxt{5}
    +\frac{g}{\Lambda v}\cfxt{6}
    +\cfgi{6} \nonumber \\
    & = -\frac{\mu^{-\epsilon}}{16\pi^2v^2}\frac{g^2 M^2}{\Lambda^2}\div.
\end{align}
\item $\Big ( \phi^\dagger \phi - \frac{v^2}{2} \Big ) (D^\mu \phi)^\dagger D_\mu \phi$
\begin{align}
 \widetilde{\lambda}_7&=  \frac{g (m^2- M^2)}{\Lambda v^3}\cfps{3}+
    \frac{g^2}{\Lambda^2v^2}( \cfps{5}+\cfps{7})+
    \frac{2 g}{\Lambda v^3}(m^2-M^2)\cfps{8}
    +\frac{g}{\Lambda v}( \cfxt{1}+
    \cfxt{4} ) \nonumber \\
    &+
    \frac{m^2-M^2}{v^2} \cfxt{3}
    +\cfgi{7} \nonumber \\
    & = - \frac{\mu^{-\epsilon}}{32 \pi^2 v^3}
    \frac{g}{\Lambda}
    \Big [ 
    M^2\Big ( 16 -14 \frac{gv}{\Lambda} + 3 \frac{g^2 v^2}{\Lambda^2} \Big ) + M_A^2 
    \Big ( 36 + 8\frac{gv}{\Lambda} - 3 \frac{g^2 v^2}{\Lambda^2} \Big )
\Big ] \div.
\end{align}
\item $F^{\mu\nu} F_{\mu\nu}$
\begin{align}
    \widetilde{\lambda}_8&\equiv\cfgi{8} = - \frac{\mu^{-\epsilon}}{96 \pi^2 v^2}M_A^2
\Big ( 2 + 2 \frac{gv}{\Lambda} + 
 \frac{g^2 v^2}{\Lambda^2} \Big ) \div .
\end{align}
\item
$\partial^\mu F_{\mu\nu} \partial^\rho F_{\rho\nu}$
\begin{align}
   \widetilde{\lambda}_9\equiv \cfgi{9}=0.
\end{align}
\item
$\Big ( 
\phi^\dagger \phi - \frac{v^2}{2}
\Big ) F_{\mu\nu}^2$
\begin{align}
    \widetilde{\lambda}_{10}&=-\frac{M^2-m^2}{v^2}\cfxt{9} + \frac{g}{v\Lambda}\cfxt{10}+\cfgi{10} \nonumber \\
    &= -\frac{\mu^{-\epsilon}}{32 \pi^2} \frac{g^2 M_A^2}{\Lambda^2 v^2 }\div.
\end{align}
\end{itemize}

\subsection{\label{sec.beta}$\beta$-functions}

It is now immediate to construct the $\beta$-functions of the theory. Renormalization implies that the running of the coupling $\widetilde{\lambda}_i$ in the target theory is determined by the corresponding $\beta$-function $\beta_i$ 
\begin{align}
	\beta_i=(4\pi)^2\frac{\mathrm{d}}{d\log\mu}\widetilde{\lambda}_i.
\end{align} 
Then, taking into accounts only terms proportional to the beyond the SM coupling $g$ we can write
\begin{align}
	\beta_i\supseteq-(4\pi)^2C_i,
\end{align}
where the coefficients $C_i$ are obtained from the corresponding $\widetilde{\lambda}_i$ dropping terms proportional to the power counting renormalizable couplings and replacing $g/\Lambda$ with $\widetilde{\lambda}_7$ as dictated by~\2eqs{tree.level}{g.invs}.

In the linear approximation we finally obtain
\begin{align}
	\beta_i\supseteq c_i\widetilde{\lambda}_7,
\end{align}  
with
\begin{align}
	c_1&=\frac1{v}(M^4-3M^4_A);& 
	c_2&=\frac2{v^3}(3M^4+M^2M^2_A-6M^4_A),\nonumber \\
	c_3&=\frac2{v^5}(5M^4+2M^2M^2_A-6M^4_A);&
	c_4&=\frac1{v}(3M^2+7M^2_A),\nonumber \\
	c_5&=0;&
	c_6&=0,\nonumber \\
	c_7&=\frac2{v^3}(4M^2+9M^2_A);&
	c_8&=\frac{1}{3v}M^2_A,\nonumber \\
	c_9&=0;&
	c_{10}&=0.
\end{align}

\section{Conclusions}\label{sec:concl}

We have presented the explicit evaluation of all the UV coefficients of dimension less or equal to 6 operators in an Abelian spontaneously broken gauge theory supplemented with a maximally power counting violating derivative interaction of dimension 6. This has been possible by following the methodology put forward in a companion paper~\cite{BQ:2019a}, in which one constructs an auxiliary theory based on the $X$-formalism in which a power-counting can be established (thus limiting the number of divergent diagrams one has to consider at each loop order) together with a mapping onto the original theory. 

In particular, a separation of the gauge-dependent contributions, associated to the cohomologically trivial invariants, from the genuine physical renormalizations of gauge invariant operators has been achieved, and we have explicitly checked in two different gauges (Feynman and Landau) our results in order to explicitly verify the gauge independence of the coefficients of gauge invariant operators. In this respect it should be clear the pivotal role played by the field redefinitions for the correct identification of the gauge dependent coefficients of the cohomologically trivial invariants and, consequently, of the coefficients of the gauge invariant operators. Purely gauge fixed on-shell calculations will completely miss their contributions, running the risk of obtaining gauge dependent results even in the case of ostensibly gauge invariant quantities. As an example we have derived the complete set of one-loop $\beta$-functions of the model which, after the field renormalization is carried out, can be read immediately from the renormalization coefficient of the corresponding operator.

The techniques presented here and in~\cite{BQ:2019a} are suitable to be generalized to: the inclusion of the complete set of dimension 6 operators; the extension of higher orders in the loop expansion; the extension of non-Abelian case, and, in particular, to the  Standard Model effective field theory in which dimension 6 operators are added to the usual SU(2)$\times$U(1) action. This latter generalization would be especially interesting, as it would allow to better understand the remarkable cancellations and regularities discovered when evaluating the one-loop anomalous dimensions for this model, and which have been linked to holomorphicity~\cite{Cheung:2015aba}, and/or remnants of embedding supersymmetry~\cite{Elias-Miro:2014eia}. Work in this direction is currently underway and we hope to report soon on this and related issues. 

\appendix

\section{List of invariants}\label{app:list}

\subsection{Pure external sources invariants}
The invariants in this sector are
\begin{align}
     &\cfps{1} \int \d  \bar c^*  ;& & 
     \cfps{2} \int \d T_1 , \nonumber \\
     &  \cfps{3} \int \d \frac{1}{2}  (\bar c^*)^2  ; 
     & & \cfps{4} \int \d  \frac{1}{2}  T_1^2  , \nonumber \\
     & \cfps{5} \int \d \frac{1}{2}  T_1\square T_1  ;
     & & \cfps{6} \int \d \frac{1}{2}  T_1\square^2 T_1  \nonumber \\
     & \cfps{7} \int \d \bar c^* T_1  ;
     & & \cfps{8} \int \d \bar c^* \square T_1  , \nonumber \\
     & \cfps{9} \int \d \frac{1}{3!}(\bar c^*)^3  ;
     & & \cfps{10} \int \d \frac{1}{2} (\bar c^*)^2 T_1 , \nonumber \\
     & \cfps{11} \int \d \frac{1}{2} (\bar c^*) T_1^2 ; 
     & & \cfps{12} \int \d \frac{1}{3!}T_1^3  .
     \label{ESinv}
\end{align}

Notice that $\cfps{6}$ has been inserted
for completeness but does not contribute
to dim. 6 operators in the target theory.

\subsection{Mixed field-external sources invariants}

The invariants in this sector are
\begin{align}
    & \cfxt{1} \int \d  \bar c^* \Big ( \phi^\dagger \phi - \frac{v^2}{2} \Big );&  
    & \cfxt{2} \int \d  T_1 \Big ( \phi^\dagger \phi - \frac{v^2}{2} \Big ),  
    \nonumber \\
    & \cfxt{3} \int \d  \bar c^* (D^\mu \phi)^\dagger D_\mu \phi;& 
    & \cfxt{4} \int \d  T_1 (D^\mu \phi)^\dagger D_\mu \phi,  \nonumber \\
    & \cfxt{5} \int \d  \bar c^* \Big [ (D^2 \phi)^\dagger  \phi + \mathrm{h.c.} \Big ];&  
    & \cfxt{6} \int \d  T_1 \Big [ (D^2 \phi)^\dagger  \phi + \mathrm{h.c.} \Big ],  \nonumber \\
    & \cfxt{7} \int \d  \bar c^* \Big ( \phi^\dagger \phi - \frac{v^2}{2} \Big )^2 ;&  
    & \cfxt{8} \int \d  T_1 \Big ( \phi^\dagger \phi - \frac{v^2}{2} \Big )^2, 
    \nonumber \\
    & \cfxt{9} \int \d  \bar c^*
    F_{\mu\nu}^2;& 
    & \cfxt{10} \int \d  T_1
    F_{\mu\nu}^2, 
    \nonumber \\
    & \cfxt{11} \int \d 
    \bar c^* T_1 \Big ( \phi^\dagger \phi - \frac{v^2}{2} \Big );& 
    & \cfxt{12} \int \d 
    T_1^2 \Big ( \phi^\dagger \phi - \frac{v^2}{2} \Big ),\nonumber \\
    & \cfxt{13} \int \d 
    (\bar c^*)^2 \Big ( \phi^\dagger \phi - \frac{v^2}{2} \Big ).&
    \label{mix.invs}
 \end{align}
Notice that the use of the contractible pair basis allows us to re-express the (otherwise present) invariants
\begin{align}
   &\cfxt{14} \int \d \bar c^* \square 
   \Big ( \phi^\dagger \phi - \frac{v^2}{2} \Big );&
   &\cfxt{15} \int \d  T_1\square 
   \Big ( \phi^\dagger \phi - \frac{v^2}{2} \Big ),
\end{align}
in terms of the above, since one has
\begin{align}
    \square \Big ( \phi^\dagger \phi - \frac{v^2}{2} \Big ) =     
   (D^2\phi)^\dagger \phi + \phi^\dagger (D^2\phi) + 2 (D^\mu \phi)^\dagger D_\mu \phi,
\end{align}
and therefore
\begin{align}
		\cfxt{14}&=2\cfxt{3}+\cfxt{5};& \cfxt{15}&=2\cfxt{4}+\cfxt{6}.
\end{align}

\subsection{Gauge invariants depending only on the fields}

The invariants in this sector are
\begin{align}
&  \cfgi{1} \int \d \Big (
    \phi^\dagger \phi - \frac{v^2}{2}
    \Big );&
& \cfgi{2} \int \d \Big (
    \phi^\dagger \phi - \frac{v^2}{2}
    \Big )^2, \nonumber \\
& \cfgi{3} \int \d \Big (
    \phi^\dagger \phi - \frac{v^2}{2}
    \Big )^3; &
& \cfgi{4} \int \d 
   (D^\mu \phi)^\dagger D_\mu \phi, \nonumber \\    
& \cfgi{5} \int \d \phi^\dagger
  [ (D^2)^2 + 
  D^\mu D^\nu D_\mu D_\nu + 
  D^\mu D^2 D_\mu ] \phi;& 
& \cfgi{6} \int \d \Big ( \phi^\dagger \phi
  - \frac{v^2}{2} \Big ) 
  \Big ( \phi^\dagger D^2 \phi + 
  (D^2\phi)^\dagger \phi \Big ),
  \nonumber \\
& \cfgi{7} \int \d 
\Big ( \phi^\dagger \phi
  - \frac{v^2}{2} \Big ) (D^\mu \phi)^\dagger D_\mu \phi; &
  & \frac{\cfgi{8}}2 \int \d F_{\mu\nu}^2 \, , \nonumber \\
 & \cfgi{9}  \int \d \partial^\mu F_{\mu\nu} \partial^\rho F_{\rho\nu}; &
  & \cfgi{10} 
   \int \d \Big ( \phi^\dagger \phi
  - \frac{v^2}{2} \Big )
F_{\mu\nu}^2.
   \  \label{g.invs}
\end{align}
These invariants are the only ones appearing also in the target theory; in that case the associated coefficient will be indicated as $\widetilde{\lambda}_i$ (with $i=1,\dots,10$).

\section{UV divergent ancestor amplitudes}
\label{app:UVdivamp}

\subsection{Tadpoles}

\begin{subequations}
\begin{align}
    \overline{\G}^{(1)}_{\bar  c^*} &= 
    - \frac{M^2+(1 - \delta_{\xi0} )M_A^2 }{16 \pi^2}\div, \label{Gc*}\\  
     \overline{\G}^{(1)}_{T_1} &= 
    - \frac{(M^4-3 M_A^4)}{16 \pi^2}\div,  \label{GT1}\\
    \overline{\G}^{(1)}_{\sigma} &=\frac{1}{16 \pi^2 v}
    \Big [ m^2 M^2 + (1 - \delta_{\xi0} ) m^2 M_A^2 + 2 (M^4 + 3 M_A^4) \Big ]\div .\label{Gs} 
\end{align}
\end{subequations}

\subsection{Two-point functions}

\begin{subequations}
\begin{align}
\overline{\Gamma}^{(1)}_{\chi^* \omega}&=\frac{eM_A^2}{8 \pi^2 v} \div (\dl-1),\label{Gch*om}\\
     \overline{\G}^{(1)}_{\chi\chi} &=  \frac{1}{32 \pi^2 v^2} \Big \{ 2 m^2 (M^2 + M_A^2) + 4 (M^4+ 3 M_A^4)
      - \frac{1}{16 \pi^2 v^2} M_A^2( m^2 + 2 p^2) \frac{\delta_{\xi0}}{\epsilon}
      \nonumber \\
    & 
    - \Big [ \frac{gv }{\Lambda} M^2 \Big ( 4 -  \frac{gv }{\Lambda} \Big ) 
    + M_A^2 \Big ( 
    8 + 12 \frac{gv}{\Lambda} +
    3 \frac{g^2v^2}{\Lambda^2} \Big ) \Big ] p^2+
    \frac{g^2 v^2}{\Lambda^2} p^4 \Big \} \div, \label{Gchch}\\
     \overline{\G}^{(1)}_{\sigma\sigma} &= \frac{1}{16 \pi^2 v^2}
     \Big \{
     2 m^4 + m^2 (5 M^2+ M_A^2) + 6 (M^4+ 3 M_A^4) \nonumber \\
     & 
     -  \Big [ 4 M_A^2 + 2 \frac{gv}{\Lambda} (m^2+ 2 M^2) 
     \Big ]
     p^2 + 
     \frac{g^2 v^2}{\Lambda^2} p^4
     \Big \}  \div 
     - \frac{ M_A^2 (m^2 + 2 p^2)}{16 \pi^2 v^2}
     \frac{\delta_{\xi0}}{\epsilon},\label{Gss}\\\
  \overline{\G}^{(1)}_{A_\mu A_\nu}  &= 
  -\frac{M_A^2}{32 \pi^2 v^2} 
  \Big \{ M^2 \frac{gv}{\Lambda} \Big (
  4 - \frac{gv}{\Lambda} \Big )  +
  M_A^2 \Big [  
  4( 4 - \delta_{\xi0} ) + 12 \frac{gv}{\Lambda}
  + 3 \frac{g^2v^2}{\Lambda^2}\nonumber \\
  &+\frac{1}{3} \Big ( 2 + \frac{gv}{\Lambda} \Big )^2 p^2
  \Big ]
  \Big \} \frac{g^{\mu\nu}}{\epsilon} 
  + \frac{M_A^2}{24 \pi^2 v^2}
  \Big ( 1 + \frac{gv}{\Lambda} + 
  \frac{g^2v^2}{\Lambda^2 }\Big )
  \frac{p^\mu p^\nu}{\epsilon}, 
  \label{GAA}\\
    \overline{\G}^{(1)}_{\bar c^* \bar c^*} &= \frac{1}{8 \pi^2}\div; \label{Gc*c*}\\
    \overline{\Gamma}^{(1)}_{\bar c^* T_1} &= \frac{1}{16 \pi^2} \Big [ 2 M^2+ 2 M_A^2 (1 - \delta_{\xi0}) - p^2 \Big ]\div,
    \label{Gc*T1}\\
    \overline{\G}^{(1)}_{T_1T_1}(p^2) &= 
    \frac{1}{32 \pi^2} \Big [ 6 (M^4 + M_A^4) - 3 (M^2+M_A^2)p^2
      + p^4 \Big ] \div,
      \label{GT1T1}\\
    \overline{\G}^{(1)}_{T_1\sigma}(p^2) &=  -\frac{1}{32 \pi^2 v}
    \Big \{ 4 m^2 (M^2 + M_A^2) + 8 (M^4 - 3 M_A^4) \nonumber \\
    & - 2 \Big ( 
    m^2 + M^2 - M_A^2 + 2 M^2 \frac{gv}{\Lambda} \Big ) p^2 +
    \frac{gv}{\Lambda} p^4
    \Big \} \div  +
    \frac{\delta_{\xi0}}{8 \pi^2 v} M_A^2 (m^2-p^2) \div, \label{GT1s}\ \\
    \overline{\G}^{(1)}_{\bar c^*\sigma}(p^2) = &
    \frac{1}{16 \pi^2 v} \Big [ -2 (m^2+M^2) + \frac{gv}{\Lambda} p^2 \Big ] \div.\label{Gc*s}
\end{align}
\end{subequations}

\subsection{Three-point functions}

\begin{subequations}
\begin{align}
    \overline{\Gamma}^{(1)}_{\bar c^* \bar c^* T_1} &= -\frac{1}{4\pi^2}\div, \label{Gc*c*T1}\\
    \left . \overline{\Gamma}^{(1)}_{\bar c^* T_1 T_1} \right |_{p_1=p_2=0} &= 
    -\frac{3 M^2 + 2 M_A^2}{8 \pi^2}\div  + \frac{M_A^2}{4 \pi^2} \frac{\delta_{\xi0}}{\epsilon}, \label{Gc*T1T1}\\ 
    \overline{\Gamma}^{(1)}_{T_1T_1T_1} &= -\frac{3 M^4}{4\pi^2}\div, \label{GT1T1T1}\\
    \overline{\G}^{(1)}_{\bar c^* T_1 \sigma} &= \frac{m^2+M^2+ \frac{M_A^2}{2} }{4 \pi^2 v}\div  
    -\frac{M_A^2}{8 \pi^2 v}\frac{\delta_{\xi0}}{\epsilon},\label{Gc*T1s}\\
    \overline{\G}^{(1)}_{\bar c^* A_\mu A_\nu}(p_1,p_2)
    &= -
    \frac{M_A^2}{16 \pi^2}\frac{g^2}{\Lambda^2} g_{\mu\nu}\div, \label{Gc*AA}\\ 
     \overline{\G}^{(1)}_{T_1 A_\mu A_\nu}(p_1,p_2)
   &= 
    \frac{M_A^2}{32 \pi^2 v^2} 
    \Big [ 
      \frac{gv}{\Lambda} \Big ( 8 - 3 \frac{gv}{\Lambda} \Big ) M^2  - 
      \Big ( 8 + 4 \delta_{\xi0} - 3 \frac{g^2v^2}{\Lambda^2}  \Big 
      ) M_A^2 \nonumber \\
      & + \frac{2}{3} \frac{gv}{\Lambda} \Big ( 1 + 2 \frac{g v}{\Lambda} \Big ) (p_1^2+p_2^2) + 2 \frac{g^2 v^2}{\Lambda^2} p_1{\cdot}p_2  \Big ]  g_{\mu\nu}\div,  \nonumber \\
      & - \frac{1}{96 \pi^2 v} \frac{g}{\Lambda} M_A^2 \Big ( 2 + \frac{gv}{\Lambda} \Big ) (p_{1\mu}p_{1\nu} + p_{2\mu}p_{2\nu})\div
      + \frac{M_A^2}{16 \pi^2}\frac{g^2}{\Lambda^2}p_{1\mu} p_{2\nu},  \label{GT1AA} \\
      \overline{\G}^{(1)}_{\sigma A_\mu A_\nu}(p_1,p_2) &=
     -\frac{M_A^2}{16 \pi^2 v^3} 
     \Big [ 
      3 \Big ( 4 + 8 \frac{gv}{\Lambda} + 3 \frac{g^2v^2}{\Lambda^2}   \Big ) M_A^2 
       - \frac{g^2v^2}{\Lambda^2} m^2 + \frac{gv}{\Lambda} \Big ( 8 - 3 \frac{gv}{\Lambda} \Big )  M^2 \nonumber \\
      &  + \frac{1}{6} \frac{gv}{\Lambda} \Big ( 4 - 10 \frac{gv}{\Lambda} +3 \frac{g^2v^2}{\Lambda^2} \Big ) (p_1^2+ p_2^2) 
      - \frac{g^2v^2}{\Lambda^2} \Big ( 4 -  \frac{gv}{\Lambda} \Big )  p_1 {\cdot}  p_2
      \Big ]
     g_{\mu\nu} \div \nonumber \\
      &      
     + \frac{1}{48 \pi^2 v^2} \frac{g}{\Lambda} M_A^2 \Big ( 2 + 7 \frac{gv}{\Lambda} \Big ) (p_{1\mu}p_{1\nu} + p_{2\mu}p_{2\nu})\div
     + \frac{M_A^2}{8 \pi^2 v} \frac{g^2}{\Lambda^2}  p_{1\mu} p_{2\nu}\div, \label{GsAA}\\
\overline{\G}^{(1)}_{\bar c^* \chi \chi} (p_1,p_2)
&= \Big [
- \frac{m^2 + M^2 - M_A^2}{8 \pi^2 v^2}  -
\frac{M_A^2}{8 \pi^2 v^2} \delta_{\xi0} 
+ \frac{1}{16 \pi^2}\frac{g}{\Lambda v} (p_1^2+ p_2^2)\nonumber \\ 
&+\frac{1}{16 \pi^2} \frac{g}{\Lambda v}
\Big ( 2 + \frac{g v}{\Lambda} \Big ) p_1 {\cdot} p_2
\Big ] \div,  \label{Gc*chch}\\
\overline{\G}^{(1)}_{T_1 \chi \chi}(p_1,p_2) 
&= \Big \{ 
    -\frac{m^2 (M^2 + M_A^2) + 2 (M^4 - 3 M_A^4)}{8 \pi^2 v^2}
    +
    \frac{m^2 M_A^2}{8 \pi^2 v^2} \delta_{\xi0}
    \nonumber \\
    &
    +\frac{1}{32 \pi^2 v^2}
    \Big [ 4 m^2 + 
    (M^2- M_A^2)
    \Big (4 + 3 
    \frac{g^2v^2}{\Lambda^2}\Big)
    + 4 M_A^2\delta_{\xi0}
    \Big ] p_1 {\cdot} p_2 \nonumber \\
    &
    + \frac{1}{16 \pi^2 v^2}
    \Big [m^2-3M_A^2+ M^2
    \Big(1+2 \frac{gv}{\Lambda} \Big )\Big ]
    (p_1^2+p_2^2)
    \Big \}\div  + {\cal O}(p^4), \label{GT1chch}\\
\left . \overline{\G}^{(1)}_{\bar c^* \sigma \sigma} \right |_{p_1=p_2=0} &= -\frac{1}{8 \pi^2 v^2} (m^2 + M^2 - M_A^2) \div 
- \frac{M_A^2}{8 \pi^2 v^2} \frac{\delta_{\xi0}}{\epsilon}, \label{Gc*ss}
\end{align}
\end{subequations}

\begin{subequations}[resume]
\begin{align}
\left . \overline{\G}^{(1)}_{T_1 \sigma \sigma} \right |_{p_1=p_2=0} &= -\frac{1}{8 \pi^2 v^2} (2 m^4 + 5 m^2 M^2+6 M^4+3m^2 M_A^2-18 M_A^4)\div  
+
\frac{3m^2 M_A^2}{8 \pi^2 v^2}
\frac{\delta_{\xi0}}{\epsilon},\\
\left . \overline{\G}^{(1)}_{\sigma T_1 T_1} \right |_{p_1=p_2=0} &= \frac{1}{8\pi^2 v}\Big[ m^2 (3 M^2 + 2 (1-\delta_{\xi 0} )M_A^2) + 6 (M^4+ M_A^4) \Big ] \div \,
,\label{GsT1T1}\\
\overline{\G}^{(1)}_{\bar c^* \bar c^* \sigma} &= 0,\label{GT1ss}\\
\left . 
\overline{\G}^{(1)}_{\sigma\sigma\sigma} 
\right |_{p_1=p_2=0}& = 
\frac{3}{8\pi^2 v^3}
\Big ( m^4 + 2 m^2 M^2+ 2 M^4 - m^2M_A^2
(1 - \delta_{\xi0}) + 6 M_A^4\Big ) \div,  \label{Gsss}\\\
\overline{\G}^{(1)}_{\sigma\chi\chi} (p_1,p_2)
& = 
\frac{1}{8\pi^2 v^3}
\Big ( m^4 + 2 m^2 M^2+ 2 M^4 - m^2M_A^2
(1 - \delta_{\xi0}) + 6 M_A^4\Big ) \div
\nonumber \\ 
& - \frac{1}{32 \pi^2 v^2}
\frac{g}{\Lambda} \Big [ 
4 m^2 + (M^2 - 3 M_A^2) \Big ( 4 + \frac{gv}{\Lambda} \Big ) 
\Big ] (p_1^2+p_2^2)
\nonumber \\
& 
-\frac{1}{16 \pi^2 v^2}
\frac{g}{\Lambda}
\Big [ 
3 \frac{gv}{\Lambda} M^2 + 
m^2 \Big ( 4 + \frac{gv}{\Lambda} \Big )
- 3 M_A^2 \Big ( 8 + \frac{3 gv}{\Lambda}
\Big ) 
\Big ] p_1 {\cdot} p_2 \div \,.\label{Gschch}
\end{align}
\end{subequations}

\subsection{Four-point functions}

\begin{subequations}
\begin{align}
     \left . \overline{\G}^{(1)}_{\sigma\sigma\chi\chi} \right |_{p_i=0}  & = 
    \frac{1}{8 \pi^2 v^4}
    \Big ( m^4+2 m^2 M^2 + 2 M^4- 2m^2 M_A^2 (1 - \delta_{\xi0} ) + 6 M_A^4 \Big ) \div, \label{Gsschch}\\
    \overline{\G}^{(1)}_{\chi\chi\chi\chi}(p_1,p_2,p_3) & =
    \frac{3}{8 \pi^2 v^4}
    \Big ( m^4 + 2 m^2 M^2 + 2 M^4 - 2 m^2 M_A^2 + 6 M_A^4 \Big ) \div \nonumber \\
    &+\frac{3}{4 \pi^2 v^4}(m^2 - M_A^2) M_A^2 \frac{\delta_{\xi0}}{\epsilon}
    -
    \frac{1}{16 \pi^2 v^3}
    \frac{g}{\Lambda}
    \Big [ 3 M_A^2 \frac{gv}{\Lambda} +
    \Big ( 8 - \frac{gv}{\Lambda} \Big )
    M^2 \nonumber \\
    &+ \Big ( 4 -\frac{gv}{\Lambda} \Big ) m^2\Big ]
    \Big ( \sum_{i=1}^3 p_i^2 + \sum_{i<j} p_i p_j \Big ) + {\cal O}(p_i^4).\label{Gchchchch}
\end{align}
\end{subequations}

\end{document}